\documentclass[structabstract]{aa}
\usepackage{graphicx}
\usepackage{txfonts}
\usepackage{natbib}
\usepackage{hyperref}
\usepackage{color}
\usepackage{subcaption}

\usepackage{amstext}

\begin{document}

   \title{Tracking the state transitions in changing-look active galactic nuclei through their polarized-light echoes}

   \titlerunning{Polarized echoes in changing-look AGNs}

   \author{F.~Marin\inst{1}\thanks{\email{frederic.marin@astro.unistra.fr}}
     \and D.~Hutsem\'ekers\inst{2}
     }

   \institute{Universit\'e de Strasbourg, CNRS, Observatoire Astronomique de Strasbourg, UMR 7550, F-67000 Strasbourg, France
     \and F.R.S.-FNRS, Institut d'Astrophysique et de G\'eophysique, Universit\'e de Li\`{e}ge, All\'ee du 6 Ao\^{u}t 19c, 4000 Li\`{e}ge, Belgium}  

   \date{Received June 18, 2019; Accepted March 13, 2020}

  \abstract
  {Variations in the mass accretion rate appear to be responsible for the rapid transitions in spectral type that are observed in 
  increasingly more active galactic nuclei (AGNs). These objects are now labeled ``changing-look'' AGNs 
  and are key objects for understanding the physics of accretion onto supermassive black holes.}
  {We aim to complement the analysis and interpretation of changing-look 
  AGNs by modeling the polarization variations that can be observed, in particular, polarized-light echoes.}
  {We built a complex and representative model of an AGN and its host galaxy and ran radiative transfer 
  simulations to obtain realistic time-dependent polarization signatures of changing-look objects. Based on 
  actual data, we allowed the system to become several times fainter or brighter within a few years, assuming a 
  rapid change in accretion rate.}
  {We obtain time-dependent polarization signatures of distant high-luminosity (quasars) and nearby 
  low-luminosity (Seyferts) changing-look AGNs for a representative set of inclinations. We predict 
  the evolution of the continuum polarization for future polarimetric campaigns with the goal
  to better understand the physics at work in these objects. We also investigate highly inclined AGNs that experience
  strong accretion rate variations without appearing to change state. We apply our modeling to Mrk~1018, the best-documented case of a changing-look AGN, and predict a variation in its polarization after the recent
  dimming of its continuum.}
  {We demonstrate that polarization monitoring campaigns that cover the transitions that are observed in changing-look AGNs 
  might bring crucial information on the geometry and composition of all the reprocessing regions within the 
  nucleus. In particular, specific features in the time variation of the 
  polarization position angle can provide a new and efficient method for determining AGN inclinations.}


\maketitle


\section{Introduction}
\label{Introduction}
The optical spectrum of type 1 active galactic nuclei (AGNs) is characterized by broad 
and narrow emission lines, while the spectrum of type 2 AGNs only show narrow emission lines. Faint broad 
H$\alpha$ emission remains detectable in type 1.9 AGN spectra, yet a faint broad H$\alpha$
emission line can alternatively mean that some scattered light is detected in the total flux, as in the case 
of 3C~234 \citep{Antonucci1984}. Changing-look AGNs (CLAGNs) constitute 
a rare class of AGNs that change their spectral type from type 1 to type 1.9/2 or vice versa on timescales
that can be shorter than a few years. CLAGNs with appearing or disappearing broad emission lines (BELs) were first
discovered among Seyfert galaxies, that is, nearby and low-luminosity AGNs \citep[e.g.,][]{Khachikian1971,Cohen1986,Goodrich1989}.
Recently, appeareance and disappeareance of BELs
was also discovered in high-luminosity AGNs (quasars) \citep{LaMassa2015,Runnoe2016,Ruan2016,MacLeod2016}. 
These spectral changes are accompanied by a dimming or a brightening of the continuum; the dimming corresponds 
to changes from type 1 to type 1.9/2, and the brighten to changes from type 1.9/2 to type 1. New CLAGNs are 
now regularly identified \citep[e.g.,][]{Gezari2017,Assef2018,Stern2018,Wang2018,Yang2018,MacLeod2019}. 

The observed changes might be caused by modifications in the source of ionizing radiation, likely a variation 
in the rate of accretion onto the supermassive black hole (SMBH): an intrinsic dimming of the continuum 
source reduces the number of photons available to ionize the gas, resulting in a net decrease of the 
emission line intensity \citep[e.g.,][]{Penston1984,Elitzur2014,Noda2018}. Alternatively, variable dust 
absorption due to the motion of individual clouds in a clumpy torus might occult the continuum source and 
the broad line region \citep[e.g.,][]{Goodrich1989,Tran1992}. The variable dust absorption scenario is 
disfavored in changing-look quasars because the timescales of extinction variations due to dusty clouds
moving in front of the broad line region are factors 2-10 too long to explain the observed spectral changes
\citep{LaMassa2015,MacLeod2016}. In addition, \citealt{MacLeod2016} have shown that if extinction by dust 
can easily explain the great change in the continuum flux and Mg~II line in several of their CLAGNS, variable 
extinction cannot explain the observed change in H$\beta$. \citet{Sheng2017} found large variations of the mid-infrared 
luminosity that echoed the optical variations that occur during the change of look. The authors argued that 
this behavior is inconsistent with the variable obscuration scenario because of the excessive amount of 
extinction needed and the too long obscuration timescale. Finally, the small continuum polarization degree measured
in most changing-look quasars also argues against variable obscuration \citep{Hutsemekers2017,Hutsemekers2019}. 
If the disappearance of BELs in changing-look quasars is indeed caused by clouds hiding the quasar core, 
only light scattered by polar material (if any; see, e.g., \citealt{Reeves2016}) can reach
the observer. The quasar light is thus expected to be highly polarized when the nucleus is
seen at not-too-low inclinations. While for nearly all changing-look \textit{\textup{quasars}} the observed changes are better
attributed to a variation in the SMBH accretion rate, by contrast, changes of look observed in \textit{\textup{Seyferts}} are 
explained by either variation in the accretion rate (e.g., Mrk1018, \citealt{Husemann2016}; Mrk590, \citealt{Denney2014}; 
NGC2617, \citealt{Shappee2014}) or variable obscuration (e.g., NGC7603, NGC2622, \citealt{Goodrich1989}; 
Mrk993, \citealt{Tran1992}). This distinction is not definitive, however, because 
objects where variable obscuration is thought to be responsible for the change of look often show complex signatures
that cannot always be explained by a change in extinction alone. For example, in the case of NGC~3516, correlated changes in 
the optical continuum and broad emission lines, strong absorption in the UV lines, as well as in the X-ray continuum
can be explained with variable obscuration \citep{Shapovalova2019}. However, shoulder-like structures in the wings
of the broad H$\alpha$ and H$\beta$ emission line profiles seem to indicate that the structure of the 
BEL region itself has significantly changed during the CLAGN activity minimum \citep{Shapovalova2019}.

In the framework of the intrinsic variability scenario, we expect a time delay between the dimming of the continuum seen
in direct light and the dimming of the scattered continuum seen in polarized light, in particular for CLAGNs 
seen at intermediate to high inclinations \citep{Hutsemekers2017,Hutsemekers2019}. The light scattered in
polar regions that extend over tens to hundreds of parsecs \citep{Capetti1995, Kishimoto2002, Zakamska2005} can
reach the observer decades to centuries after the direct light in the most extreme cases. Fortunately, light echoes most 
likely originate from the base of the wind \citep{Marin2012}, where shorter (months to years) time delays are expected.
In this case, the polarized light contains the echo of a past bright phase diluted by the much fainter direct light,
then resulting in a high polarization degree that is expected to slowly decrease with time.  \citet{Hutsemekers2019} 
showed that such a scenario is supported by time-dependent radiative transfer simulations, and argued that it can 
explain the high polarization measured in the changing-look quasar J022652.24$-$003916.5. BELs are also expected to be observed
in the polarized light, constituting echoes of a past type 1 phase.

In the present paper, we further explore the effect of an intrinsically variable continuum source on AGN 
polarization. Based on time-dependent radiative transfer simulations, we produce polarization time series 
for various realistic CLAGN models. In Sect.~\ref{Model} we describe the radiative transfer code and the 
generic AGN model we use to obtain time-dependent polarization signatures of CLAGNs. We investigate both 
dimming and brightening scenarios for quasars and Seyfert-like objects seen at various inclinations. 
Using the best-documented observational case, we apply our simulations to Mrk~1018 in Sect.~\ref{Application}
to show how the polarization signatures are expected to vary within the next decades. We discuss the 
constraints that future polarimetric monitoring campaigns of known CLAGNs might advance our understanding of the 
morphology and composition of AGNs, and examine the most promising cases in Sect.~\ref{Discussion}. We 
finally conclude our analysis in Sect.~\ref{Conclusions} by summarizing the benefits of regular polarimetric 
observations of AGNs and the importance of spectropolarimetric observations in addition to 
single broadband-imaging polarimetric measurements.

\section{Modeling polarized echoes}
\label{Model}
We start our investigation by constructing a generic AGN model in Sect.~\ref{Model:QSO}. We explore 
its polarization properties from a static point of view in Sect.~\ref{Model:Static}, that is, when the AGN does not experience 
a strong change in its optical classification. We then allow the model to 
evolve by dimming or brightening its nuclear flux by a factor 10 in 6 years
in Sect.~\ref{Model:Dynamic}. This corresponds to the timescale of the observed BEL disappearance in the 
quasar J022652.24$-$003916.5 \citep{MacLeod2016,Hutsemekers2019}.

\subsection{ AGN model}
\label{Model:QSO}

\begin{figure}
  \centering
  \includegraphics[trim = 0mm 0mm 0mm 0mm, clip, width=9.0cm]{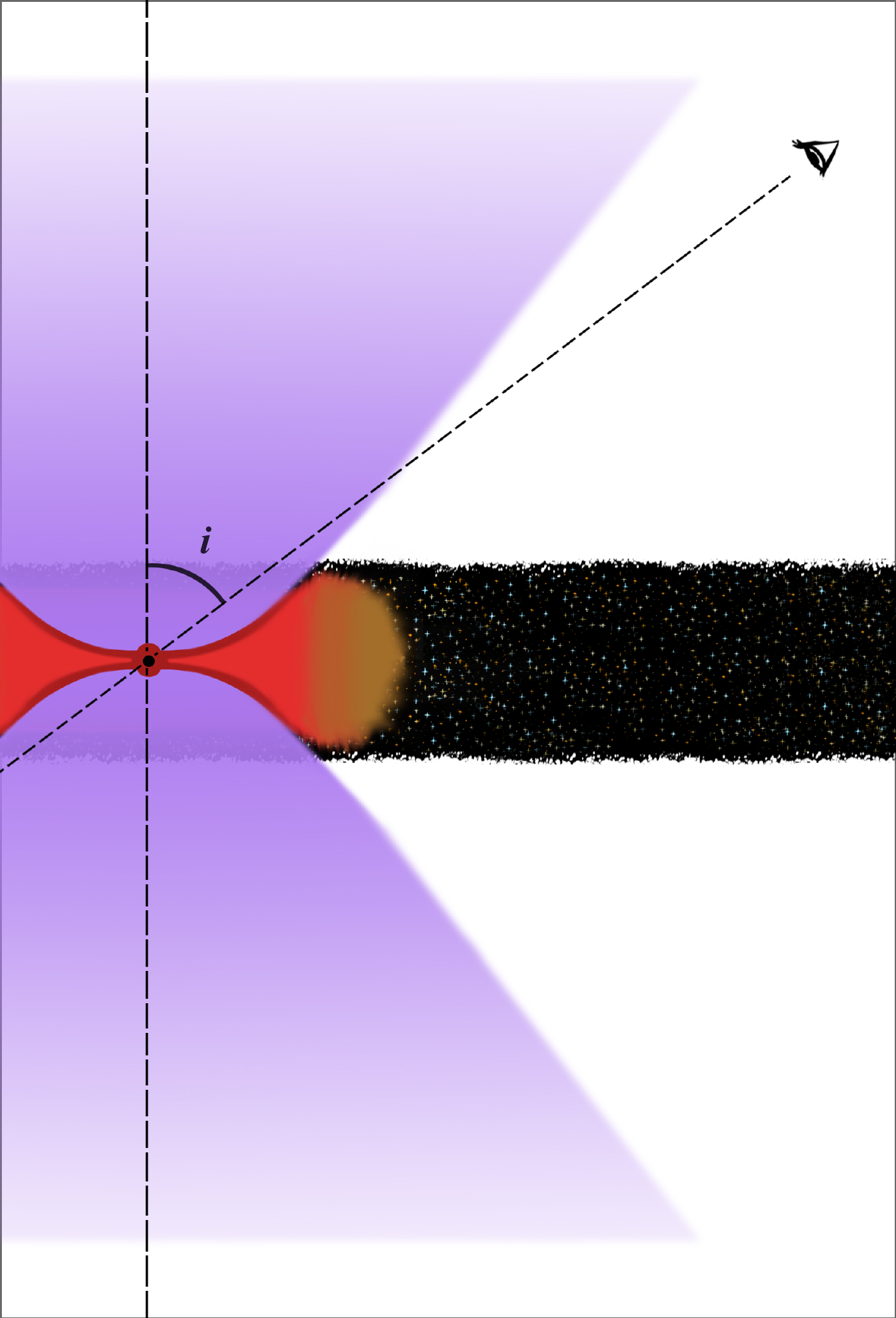}
  \caption{Unscaled illustration of the AGN model components.
          The central SMBH and its accretion disk are
          merged into a punctual source represented
          with a black dot.  By moving away from the SMBH, the source becomes the 
          electron-scattering region (red). The equatorial 
          electron scattering region becomes the 
          circumnuclear dusty torus (brown). The 
          inner dusty funnel collimates disk-born polar
          winds (violet). Finally, the host galaxy 
          is shown in black. The vertical long-dashed 
          line is the symmetry axis of the AGN. The 
          short-dashed line is the observer's viewing
          angle, whose inclination is defined with 
          respect to the symmetry axis of the model. 
          See text for further details.}
  \label{Fig:Scheme}
\end{figure}

We built a radio-quiet AGN prototype on the basis of the unified model \citep{Antonucci1993}. At 
the center of the simulation lies an SMBH and its accreting and emitting engine, which we modeled
using a point-like source that emits unpolarized photons isotropically. In the remainder of this article, we 
focus on mono-energetic analyses at UV-visible wavelengths, therefore the exact spectral energy distribution does not 
matter. Around the central engine, we added an equatorial, electron-filled, uniform flared-disk component that is
necessary to explain the temporal variations in the polarization degree of the H$\alpha$ and
continuum emission, as well as the polarization position angle of the broad H$\alpha$ wings, as observed in nearby AGNs \citep{Young2000}.
Because the variations observed in the polarized flux are slower than the observed variations in the direct view 
of the broad lines \citep{Young1999}, this scattering region must be larger than the BEL and extend up to the 
torus. The half-opening angle of the flared disk was set to 20$^\circ$ from the equatorial plane and its optical 
thickness in the V-band was set to $\tau_V$ = 1. Both values were derived from the extensive grid of models presented 
in \citet{Marin2012}. They allow us to reproduce the expected optical polarization signatures of nearby 
Seyfert galaxies. The inner and outer radii of this scattering region were set to 0.001 and 0.1~pc, respectively.
The inner radius corresponds to a distance large enough so that the central engine can be approximated 
by a point source, in agreement with self-consistent dynamical models of the BEL region \citep{Czerny2017}.
The outer radius was bounded by the dust sublimation radius that defines the beginning of the circumnuclear region, which is usually 
referred to as the torus. Its inner radius is luminosity dependent and scales as 
R$_{\rm in} \approx$ 0.4 L$_{\rm 45} ^{1/2}$ T$_{\rm 1500} ^{2.6}$ pc \citep{Nenkova2008a,Nenkova2008b}. For a fiducial
dust sublimation temperature T = 1500K and a bolometric luminosity L = 10$^{45}$ erg.s$^{-1}$, the inner radius 
of the torus is a fraction of a parsec, under the hypothesis of an optically thick dusty region and for a specific 
composite dust grain structure \citep{Elitzur2008}. We modeled the torus using a flared-disk region, and we filled  
it with dust grains (37.5\% graphite and 62.5\% astronomical silicate). The grain radii ranged from 0.005 to 
0.250 $\mu$m and the grain size index was set to -3.5. The model was the usual cosmic dust of the Milky Way 
\citep{Mathis1977} and the optical thickness of this uniformly filled medium was in excess of 50, so that 
the torus was optically thick to ultraviolet and optical radiation. The half-opening angle of the torus
was set from observational constraints based on quasar populations \citep{Sazonov2015,Marin2016} 
and fixed at 45$^\circ$. Its outer radius was fixed by mid-infrared interferometric measurements in 
nearby AGN, together with dust radiative transfer simulations that all point toward a rather 
compact torus (R$_{\rm out} \sim$ 5 -- 10~pc) embedded in a larger diffuse dusty region 
\citep{Pier1992,Meisenheimer2008}. We therefore set this outer radius to 5~pc. The dusty structure collimates 
the winds from the central engine. They take the form a biconical structure that extends 
up to 100~pc along the polar direction, the so-called ionization cones resolved in nearby Seyferts 
\citep{Antonucci1994,Simpson1997}. The optical depth of the medium was chosen to be 0.1 so 
that a non-negligible fraction of the photons can scatter inside the polar medium \citep{Marin2012}.
Finally, around the AGN model, we added a geometrically thin disk-like structure representative 
of the host galaxy. It was assumed to be perpendicular to the AGN symmetry axis and to extend up 
to several kiloparsecs. This is not the most general case, however, because it is known that 
the position angles of AGNs and their host are uncorrelated \citep{Ulvestad1984,Kinney2000}.
This choice is due to limitations of our simulation tool. The host galaxy emits unpolarized starlight 
that dilutes the observed polarization from the central engine. Absorption and/or 
scattering by dust in the host were not taken into account. By doing so, we mitigated the effect of the 
aligned AGN/host configuration by preventing scattering-induced or dichroic absorption polarization from 
the host galaxy \citep{Thompson1988}. In addition, because we work under the approximation of single-wavelength photons,
reddening of the type 1 spectra by material in the host plane is not a concern either. In our simulations, the 
amount of starlight is fixed; only the nuclear light from the AGN can dim or brighten due to mass accretion
rate variations \citep{Noda2018}. An unscaled sketch of the AGN model is presented in Fig.~\ref{Fig:Scheme}
to illustrate the complexity of the simulations that coherently couple emitting, absorbing, and reprocessing
regions from milliparsec to kiloparsec scales.

\subsection{Static polarization properties}
\label{Model:Static}

\begin{figure}
  \centering
  \includegraphics[trim = 0mm 0mm 0mm 1mm, clip, width=9.4cm]{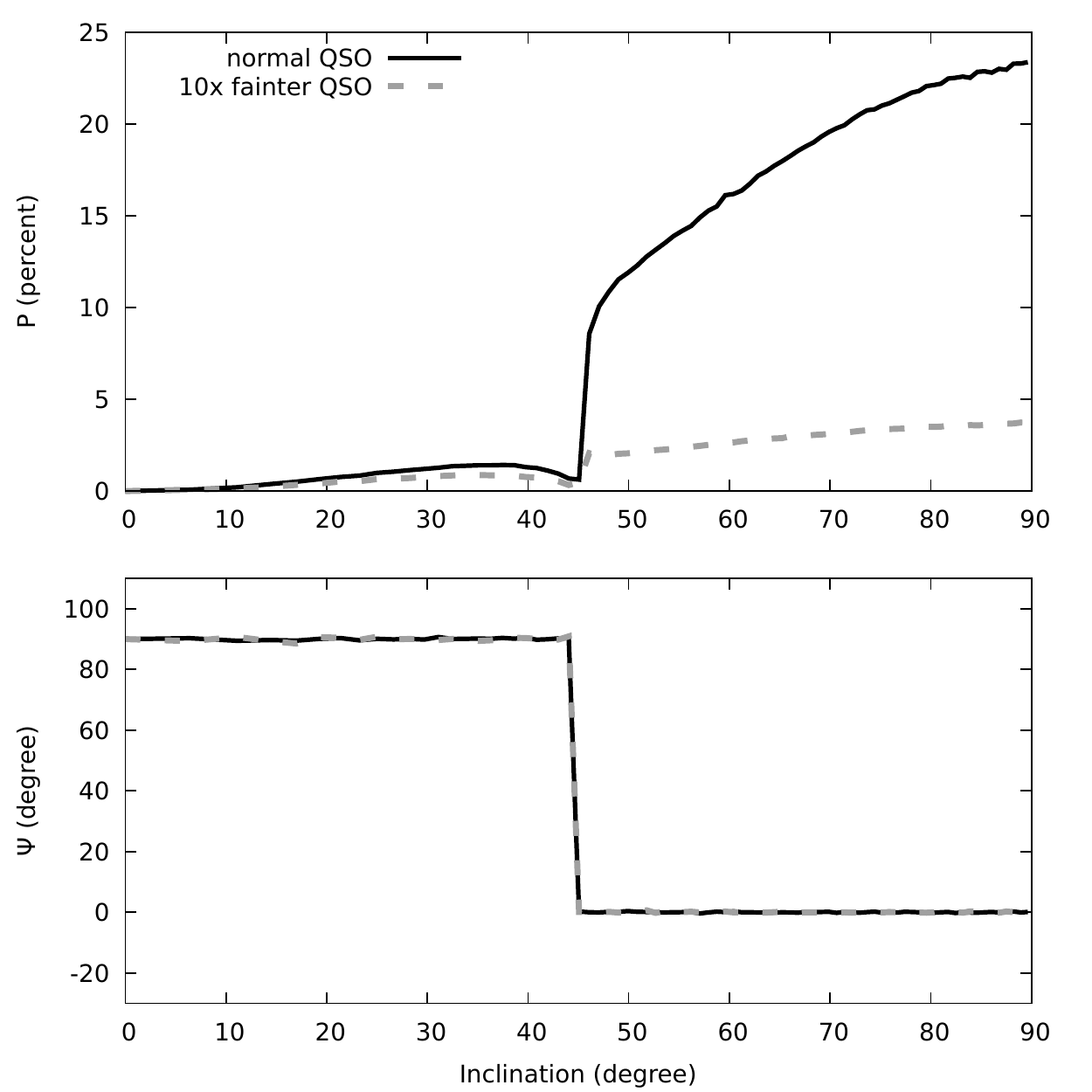}
  \caption{Polarization degree (top) and polarization 
    position angle (bottom) from our AGN model.
          The black 
          line corresponds to a bright
          quasar. In this case, the host galaxy contributes
          $\sim$ 5\% of the total light along type 1 inclinations ($i \leq 45 ^\circ$).
          The long-dashed gray line 
          corresponds to the exact same quasar, but ten 
          times fainter.
          The small fluctuations in $P$ 
          are due to statistical noise.}
  \label{Fig:Pol_vs_i}
\end{figure}

To simulate the radiative transfer of photons from the SMBH and its accretion disk
to the observer, we used the Monte Carlo code {\sc stokes} \citep{Goosmann2007,Marin2012,Marin2015,Rojas2018,Marin2018}.
The code allows us to reproduce the radiative coupling of photons that travel within a complex 
three-dimensional environment. All scattering physical processes that are relevant in the optical regime (Thomson, 
Mie, Rayleigh) are included in the version of the code we used, and were used when the spectropolarimetric
signal of CLAGNs was considered. The code is available online and can 
be used freely: \url{http://stokes-program.info}. Photons that successfully escaped the model are 
registered by a virtual web of detectors in all polar and azimuthal angles at once. The Stokes parameters 
of light allow us to reconstruct the observed polarization degree $P$ and polarization position angle $\Psi$.
In the following we denote with $\Psi$ the polarization position angle defined with respect to the AGN symmetry axis, 
$\Psi$ = 90$^\circ$ indicating polarization parallel to the axis. We denote with $\theta$ the polarization position 
angle measured  with respect to the equatorial north-south direction,  $\theta$ = 90$^\circ$ corresponding to the 
east-west direction. For the remainder of this paper, we concentrate our efforts on the continuum polarization, and we constrain the source
to emit at a central wavelength of 2900~\AA. This value is derived from the recent observations of changing-look
quasar polarization by \citet{Hutsemekers2019}, who measured the continuum polarization of J022652.24$-$003916.5
at 2900~\AA~(in the quasar rest frame). Additionally, in the near-ultraviolet band, the amount of starlight 
diminishes, which increases our chances of observing the true nuclear polarization signal. We simulated 10$^9$
photons for each realization and azimuthally integrated the signal due to the axisymmetric nature of the model.
With the exception of the results presented in Fig.~\ref{Fig:Pol_vs_i} that took only 7 hours to complete, 
the simulations took $\sim$ 50 hours each on a regular desktop computer.

In Fig.~\ref{Fig:Pol_vs_i} we explore the inclination-dependent polarization degree (top) and angle 
(bottom) from our AGN model (solid black line). The contribution of the host galaxy to the total flux 
was set to 5\% (typical value for quasars; see, e.g., \citealt{Kauffmann2003}). The polarization 
degree increases with increasing inclination. This is due to the departure of the centro-symmetric 
pattern of polarization vectors seen at $i=$ 0$^\circ$. The polarization degree reaches 1.4\% at most 
around $i=$ 35$^\circ$. The associated polarization position angle is 90$^\circ$, which corresponds to a 
polarization angle parallel to the symmetry axis of the model according to the convention used in 
{\sc stokes}.
This is the expected polarization angle observed in both type 1 quasars and Seyfert 
galaxies \citep{Antonucci1993}. When the observer's line of sight reaches the geometrical limits of 
the circumnuclear dusty region, there is a canceling action from parallel $\Psi$ emerging from the 
equatorial regions and perpendicular $\Psi$ originating from the polar regions. The polarization 
degree decreases and becomes null around $i=$ 45$^\circ$. Then $P$ rises again with inclination and $\Psi$ 
becomes perpendicular ($\Psi$ = 0$^\circ$) because the central engine is now 
obscured by the optically thick dusty layer and the photons we observe have been scattered inside the
polar winds, resulting in higher polarization degrees (up to 23\%). High polarization degrees 
like this are common after correction for the stellar light \citep[e.g.,][]{Kay1994}, and some broad line 
polarizations or some spatially resolved regions of NGC~1068 have even higher values \citep{Antonucci2002,Simpson2002}.

We then decreased the contribution from the central engine to the total flux to estimate the polarization 
we would observe from a quasar that is ten times fainter. This corresponds to the Seyfert galaxy case where 
the host or AGN flux ratio has increased by a factor 10. We caution that in this example, the model 
is static: the flux did not dim from a bright state, and we instead considered the AGN to be naturally less 
bright. Fig.~\ref{Fig:Pol_vs_i} (dashed gray line) shows that a fainter AGN presents a slightly 
lower polarization degree along type 1 viewing angles (between 0$^\circ$ and 45$^\circ$). On average, the 
polarization is 1.5 times lower because the starlight dilution is stronger. Along type 2 
inclinations ($i=$ 45$^\circ$ -- 90$^\circ$), the difference is much more notable, with a variation by a factor
$\sim$ 6 because the central engine is only seen through reprocessed emission 
that is ultimately heavily diluted by the host emission. We note, however, that the polarization position
angle does not change between a quasar and its ten times fainter counterpart.

\subsection{Polarized echoes of CLAGNs}
\label{Model:Dynamic}

We now investigate the time-dependent variations in polarization that result from
a dimming or brightening of the central AGN. We use the exact same model
investigated in Sect.~\ref{Model:Static} and focus on the most luminous objects
(quasars) for observational reasons: their host or AGN flux ratio is much more
favorable to the detection of linear continuum polarization in the
near-ultraviolet and optical wavelength bands. First, we set the quasar to its
bright state for 50 years, then allow a linear decrease in quasar flux by a
factor 10 in 6 years. We then track the evolution of polarization during the
next century in Sect.~\ref{Model:Dynamic:Dimming}. Second, in
Sect.~\ref{Model:Dynamic:Brightening}, we use the faint quasar model and turn on
the central engine to determine whether the polarization behavior is symmetric to
the dimming case or if it has unique signatures. Finally, we compare quasar and Seyfert changing-look AGNs in
Sect.~\ref{Model:Dynamic:Seyferts}.

\subsubsection{Dimming quasars}
\label{Model:Dynamic:Dimming}

\begin{figure*}
    \centering
    \begin{subfigure}[b]{0.475\textwidth}
        \centering
        \includegraphics[width=\textwidth]{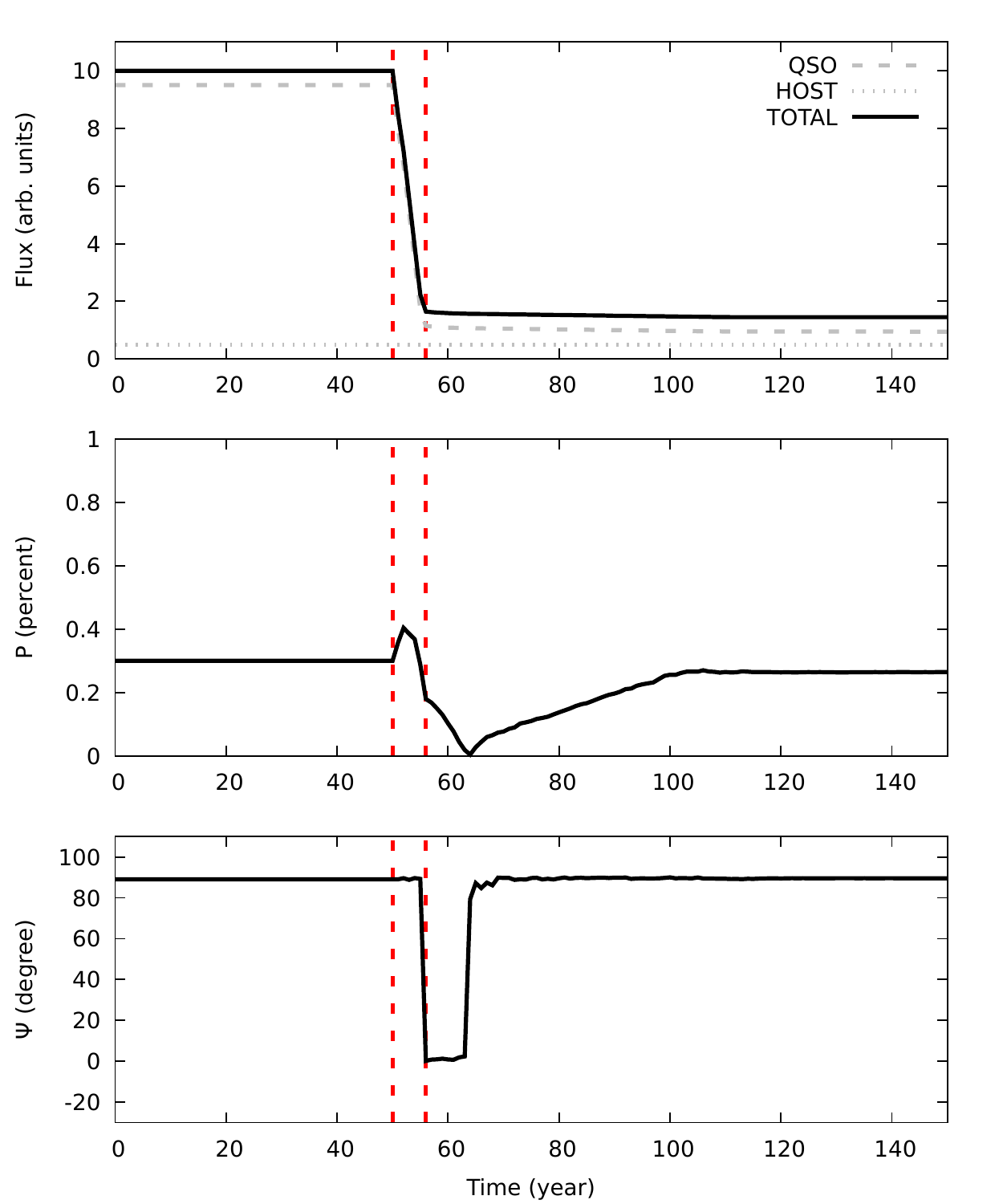}
        \caption{\small Observer inclination $\sim$ 10$^\circ$ 
          from the polar axis of the model.}    
        \label{Fig:10deg}
    \end{subfigure}
    \hfill
    \begin{subfigure}[b]{0.475\textwidth}  
        \centering 
        \includegraphics[width=\textwidth]{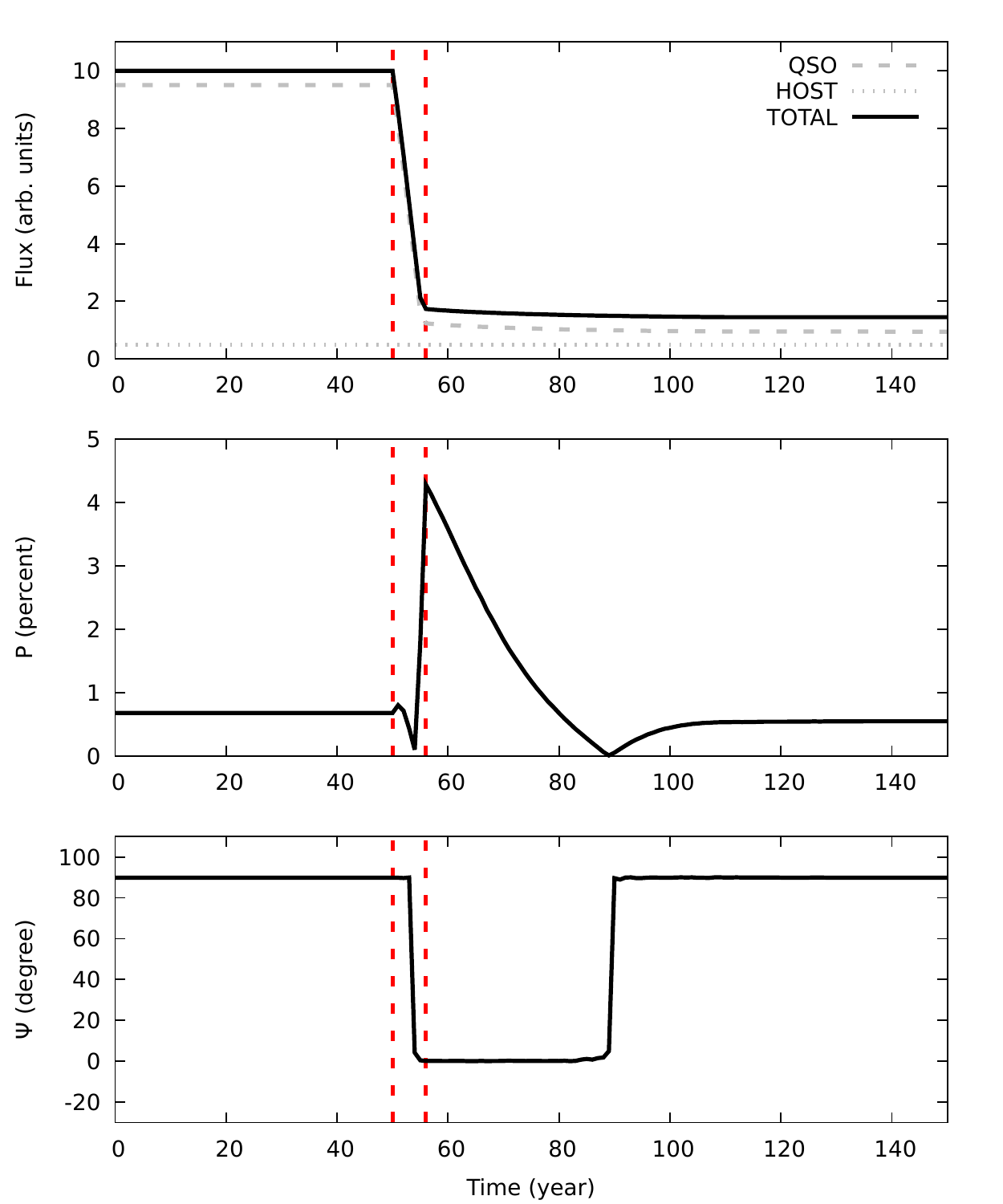}
        \caption{\small Observer inclination $\sim$ 45$^\circ$ 
          from the polar axis of the model.}     
        \label{Fig:45deg}
    \end{subfigure}
    \vskip\baselineskip
    \begin{subfigure}[b]{0.475\textwidth}   
        \centering 
        \includegraphics[width=\textwidth]{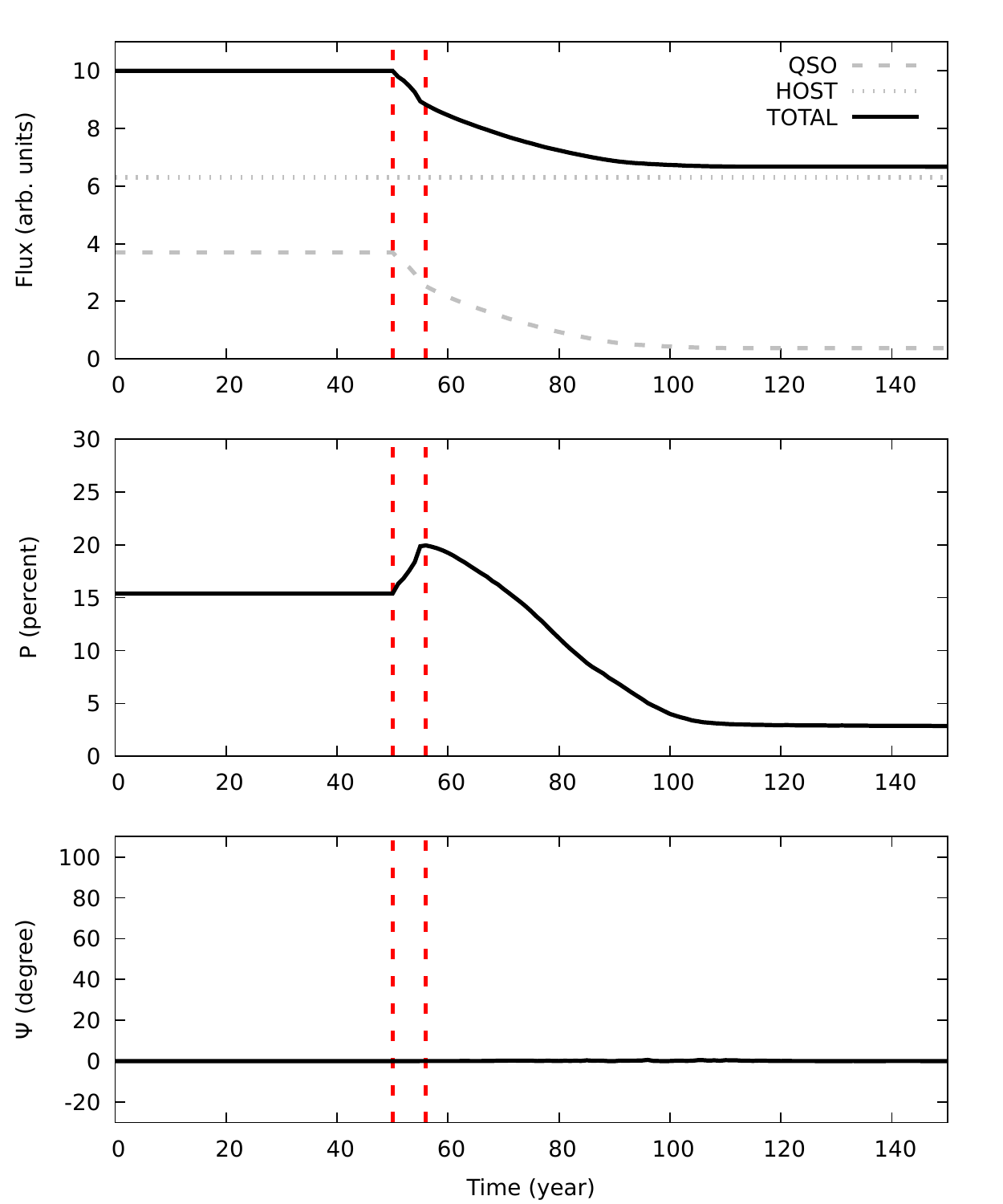}
        \caption{\small Observer inclination $\sim$ 60$^\circ$ 
          from the polar axis of the model.}    
        \label{Fig:60deg}
    \end{subfigure}
    \quad
    \begin{subfigure}[b]{0.475\textwidth}   
        \centering 
        \includegraphics[width=\textwidth]{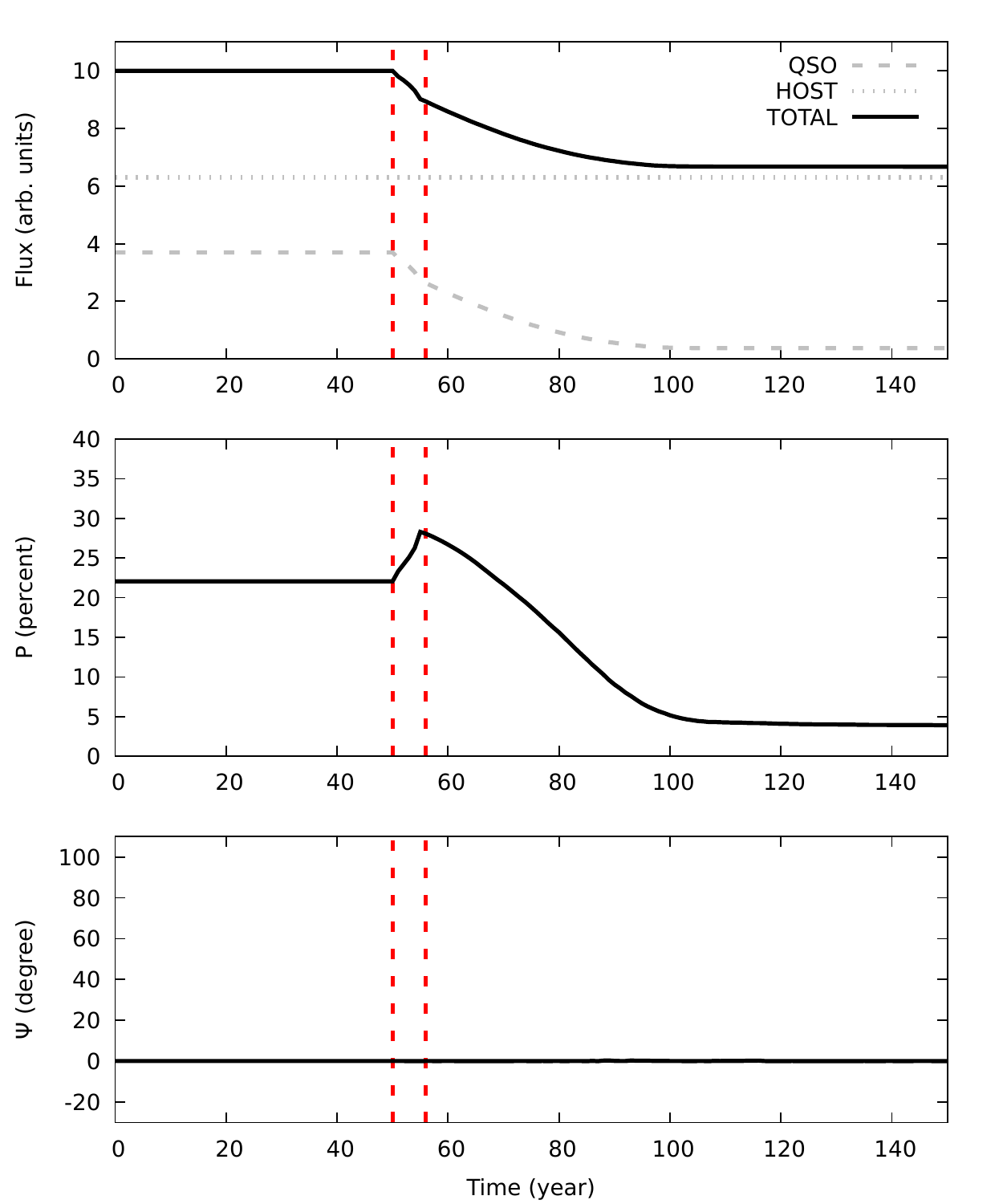}
        \caption{\small Observer inclination $\sim$ 80$^\circ$ 
          from the polar axis of the model.}  
        \label{Fig:80deg}
    \end{subfigure}
    \caption{Variation in normalized total flux (top panel), 
          polarization degree (middle), and polarization 
          position angle (bottom) from our changing-look 
          quasar model seen along four different viewing angles: 
          10$^\circ$ (\textit{a}), 45$^\circ$ (\textit{b}), 
          60$^\circ$ (\textit{c}), and 80$^\circ$ (\textit{d}).  
          In the total flux panels, we show the contribution
          of the quasar, the host galaxy, and the 
          combination of both using gray dashed lines 
          for the first two angles and a black solid line
          for the other two angles. Between the first and second 
          long-dashed vertical red lines, the central 
          engine dimmed by a factor of 10 in 6 years.} 
    \label{Fig:QSO_dimming}
\end{figure*}

We present the results of our polarimetric investigations of dimming quasars in Figs.~\ref{Fig:10deg}, \ref{Fig:45deg}, \ref{Fig:60deg},
and \ref{Fig:80deg}. The first two panels show the time-dependent signal from type 1 quasars, that is, objects seen at an inclination
of 10$^\circ$ and 45$^\circ$, respectively. In both cases the central engine is directly visible through the equatorial dust funnel,
although in the second case the observer's line of sight grazes the dust horizon. The other two panels 
show the polarization expected from type 2 quasars, that is, when the extinction of the nucleus is total, at inclinations 60$^\circ$ and 
80$^\circ$, respectively. The panels show the total flux variation (top panel), polarization degree (middle), and polarization position
angle (bottom) as a function of time. In the total flux panel, we show the contribution of the quasar, the host galaxy, and the combination
of both. We indicate the onset and end of the dimming phase using vertical red lines. 

The type 1 objects (Figs.~\ref{Fig:10deg} and \ref{Fig:45deg}) show that the flux variation is rather straightforward. 
The flux diminishes by a factor close to but not exactly 10 in 6 years (the simulation in Fig.~\ref{Fig:45deg} was shown 
in \citet{Hutsemekers2019}, where the quoted five-year dimming corresponds to the bulk of the flux drop). This is due to the contribution
of the host galaxy, which remains constant with time. During the bright quasar state, the host only accounts for 5\% of the total flux 
from the system, but when the central engine is in the low state, the relative contribution of the host is higher, almost 30\%. After 
the tenfold decay, the AGN optical type classification can change from type 1 to true type 2 (i.e., intrinsically lacking a BEL emitting
region) Note that it is still debated whether true type 2s, intrinsically lacking broad line emission and in some cases, a 
thermal big blue bump truly exist \citep{Antonucci2012}. One of the best representative case of true type-2s, NGC~3147, has recently 
revealed an intrinsic H$\alpha$ line with an extremely broad base (full width at zero-intensity on the order of 27\,000~km.s$^{-1}$), 
see \citet{Bianchi2019}. The variation in type classification is rather straightforward to track using medium-resolution spectroscopic observations (e.g., 
\citealt{Cohen1986,LaMassa2015}). The polarization resulting from the state change, however, is more complex to predict and observe. It is rather different between the two inclinations.

In the 10$^\circ$ case (Fig.~\ref{Fig:10deg}), the polarization degree 
is 0.3\%, with a parallel ($\Psi$ = 90$^\circ$) polarization angle during the bright state. The low polarization is due to the almost
face-on inclination of the nucleus. Starlight dilution plays a minor role here. When the flux starts to dim, the polarization degree 
increases to 0.4\% for a very short time ($\sim$ one year), then decreases sharpl to zero in a decade. The increase in 
polarization is due to the combined action of 1) a lower amount of direct unpolarized flux from the central engine, and 2) a
constant amount of reprocessed (delayed) radiation from the equatorial region. The reprocessed radiation is an echo from the past 
bright state. Its flux is thus higher and outshines the constant weaker unpolarized flux from the host, explaining the increase in polarization. The duration of the high-polarization peak depends on the distance of the scatterer from the source and can be used to 
achieve polarized reverberation mapping of the inner quasar regions, such as has been demonstrated by \citet{Gaskell2012}, \citet{Afanasiev2014} and \citet{Afanasiev2015}.
When the echo has faded away, the smaller number of source photons that is scattered inside the equatorial structures results in a decrease 
in polarization degree. At a certain point, the polarization position angle rotates and becomes perpendicular ($\Psi$ = 0$^\circ$) 
for less than a decade. This is due to a second polarized echo coming from the extended polar region. The ionization cones are indeed much
more extended than the equatorial scattering region and have a lower electron density. The polarized echo thus takes longer 
to be visible. The number of photons from the past bright quasar state that have scattered inside the extended polar regions becomes larger
than the ernumb of photons from the dimmed source that is seen without additional time delay. Because scattering occurs in the polar 
region, the polarization degree rotates and becomes perpendicular. When this secondary polarized echo has faded away, $\Psi$ rotates
again as a result of the contribution of the equatorially scattered flux. The orthogonal flip of the polarization angle is responsible for the 
cancellation of the polarization degree, which can slowly rise again afterward. The final linear polarization degree, after 
the quasar has dimmed and polarized-light echoes have vanished, is about 0.27\%. This value is lower than during the bright state 
because fraction of starlight from the host is higher. It takes about 40 years for the light echoes to stop propagating inside the 
model, showing that polarization can track down changing-look objects for a much longer period than photometry alone. Unfortunately, 
the low levels of linear polarization degrees at low inclinations makes the variation in $P$ difficult to measure because observational
errors on $P$ are commonly about 0.1\%. To facilitate polarimetric monitoring campaigns, it is therefore better to study the 
polarization of AGNs seen at higher inclination that represents still type 1s.

The variation in polarization in the case of the highest possible
type 1 inclination (45$^\circ$, Fig.~\ref{Fig:10deg}) shows a different behavior that is associated with longer timescales.
The polarization degree first increases by a negligible amount as a result of the polarized-light echo from the BEL or scattering region, similarly to the 10$^\circ$
inclination case. Then $P$ sharply decreases as the polarization position angle rotates from parallel to perpendicular. This is due to 
scattering of photons from the base of the polar regions. The higher system inclination promotes higher polarization degrees ($P$ induced 
by Thomson laws is cosine-squared-dependent on the scattering angle). Thus the polarization degree is dominated by a polar component
and is about 4 -- 5\%. Then, similarly to lower inclinations, the polarized-light echo fades away with a longer timescale because of 
the higher inclination (this is a geometric effect). This tells us that the transitory phase of the polarization position angle might be related 
to the inclination angle, allowing an independent and novel way of measuring quasar inclination. Highly inclined type 1 quasars are thus
interesting targets because we can expect higher observable polarization degrees and strong time-dependent polarization signatures. 

In the case of type 2 quasars (Figs.~\ref{Fig:60deg} and \ref{Fig:80deg}), the situation is different but particularly interesting
because we investigate a transition from a type 2, that is, a  hidden or obscured type 1, to a true type 2 (i.e., intrinsically lacking a BEL 
region). Type 2 objects are often more complex to observe in polarization because starlight dilution is strong. The flux from 
the central quasar regions is obscured by the circumnuclear dusty component and can only be detected through reprocessed photons in
the polar cones \citep{Miller1983,Antonucci1985}. This results in stronger host flux contributions, which commonly exceed 50\% in the case 
of type 2 quasars (see, e.g., \citealt{Zakamska2006}). Because of the geometric configuration and optical depth of the various 
reprocessing regions in our model, only 3\% of the quasar continuum flux can be detected in type 2 inclinations through polar scattering.
As a consequence, Figs.~\ref{Fig:60deg} and \ref{Fig:80deg} (top panels) show that during the bright state, 37\% of 
the observed total flux is due to the AGN and 67\% is due to the host. The host contribution
becomes preponderant, up to 94\%, after the quasar has dimmed by a factor of 10. This results in a total flux attenuation that is only a 
factor 1.5 lower after the change of look. The transition also takes longer, many decades, because we only see light echoes in the extended
polar region. These echoes naturally carry a high polarization degree. In the case of a 60$^\circ$ inclination, $P$ = 15.4\% in the 
bright state. For a 80$^\circ$ inclination, $P$ = 22.1\%. Both inclinations show a perpendicular polarization angle, the usual signature 
of polar scattering. When the central engine starts to dim, $P$ increases by a few percent, then slowly decreases with time, at the same pace 
as the total flux. The polarization position angle remains constant over time. Because the total and polarized fluxes are only due
to polar scattering, these were expected behaviors. The short increase in $P$ is due to the radiative coupling of the various quasar 
components. Most of the polarized flux we observe is due to polar scattering in the ionization cones. However, a smaller fraction is also 
due to scattering of photons inside the equatorial dust funnel. The inclination of the quasar means that a small fraction of radiation can
escape by (back)scattering from the dust wall opposite to the observer's side \citep{Marin2012}. This low photon flux carries a parallel
polarization that weakly dilutes the final perpendicular polarization we observe. When the central engine dims, this equatorial contribution
decreases more rapidly than the polar echo because of the different spatial extensions of the components. This results in less dilution, 
hence a marginally higher polarization degree. When the quasar has dimmed by a factor 10, the system stabilizes and the polarization 
degree decreases with time. The final polarization we observe is much lower than in the bright phase because the contribution 
from host starlight is higher. In the 80$^\circ$ inclination case, $P$ ends at 3.9\% (Fig.~\ref{Fig:80deg}, middle panel). Our study of changing-look quasars with inclinations higher than 45$^\circ$ demonstrates that the  transition of a type 2, that is, a hidden or obscured 
type 1 to a true type 2 can be detected with polarimetry.

\subsubsection{Brightening quasars}
\label{Model:Dynamic:Brightening}

\begin{figure*}
    \centering
    \begin{subfigure}[b]{0.475\textwidth}
        \centering
        \includegraphics[width=\textwidth]{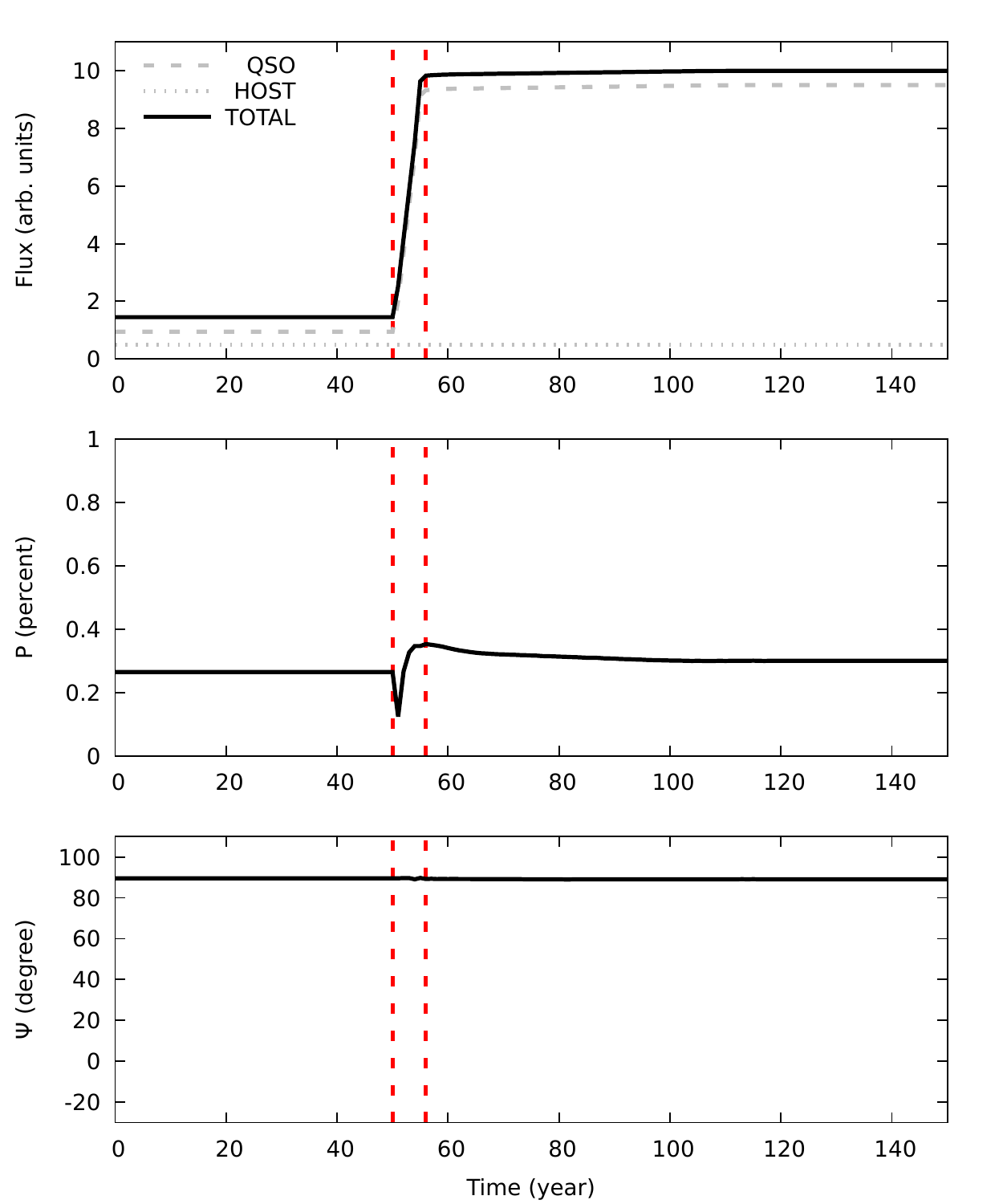}
        \caption{\small Observer inclination $\sim$ 10$^\circ$ 
          from the polar axis of the model.}    
        \label{Fig:10deg_brightening}
    \end{subfigure}
    \hfill
    \begin{subfigure}[b]{0.475\textwidth}  
        \centering 
        \includegraphics[width=\textwidth]{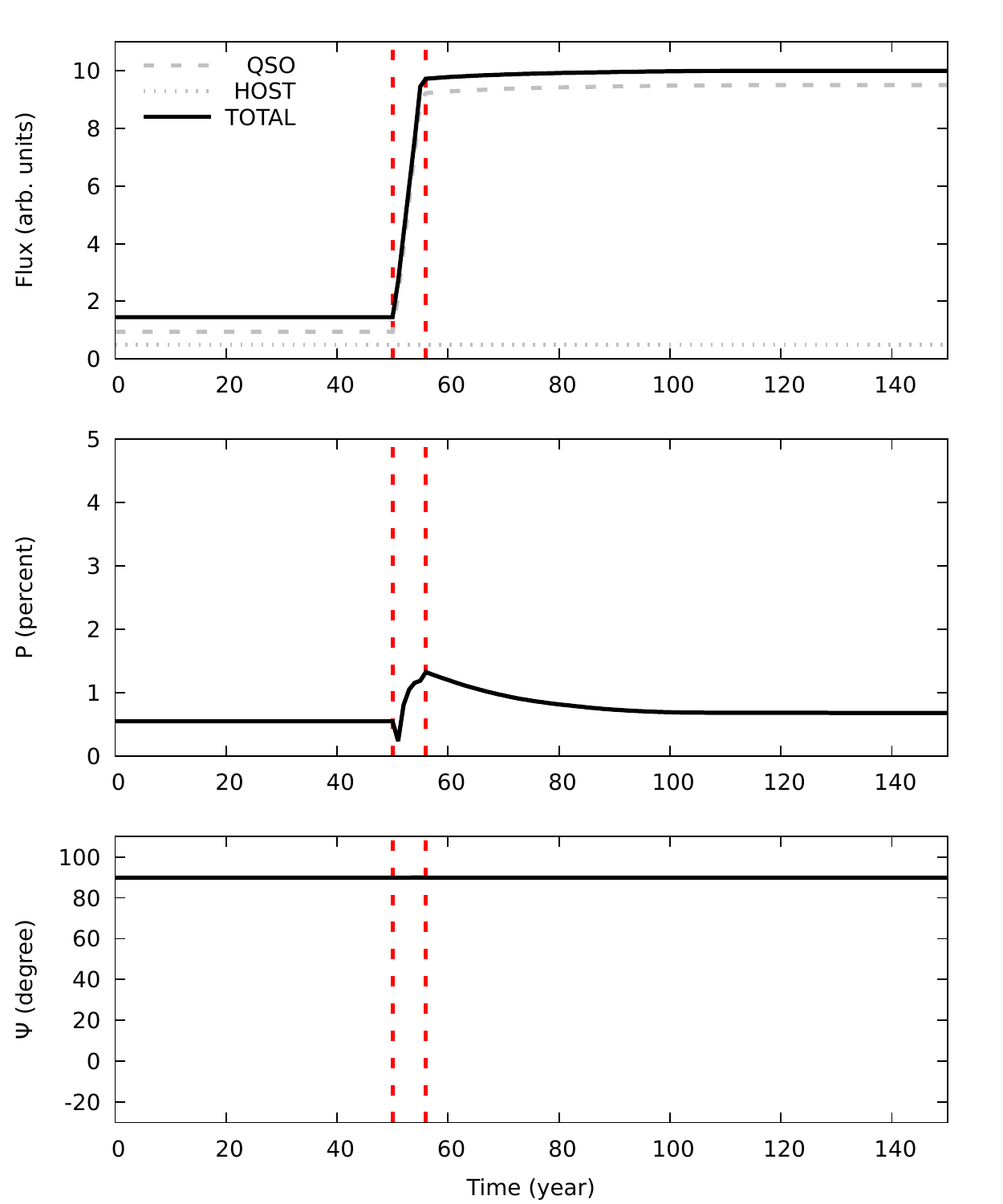}
        \caption{\small Observer inclination $\sim$ 45$^\circ$ 
          from the polar axis of the model.}     
        \label{Fig:45deg_brightening}
    \end{subfigure}
    \vskip\baselineskip
    \begin{subfigure}[b]{0.475\textwidth}   
        \centering 
        \includegraphics[width=\textwidth]{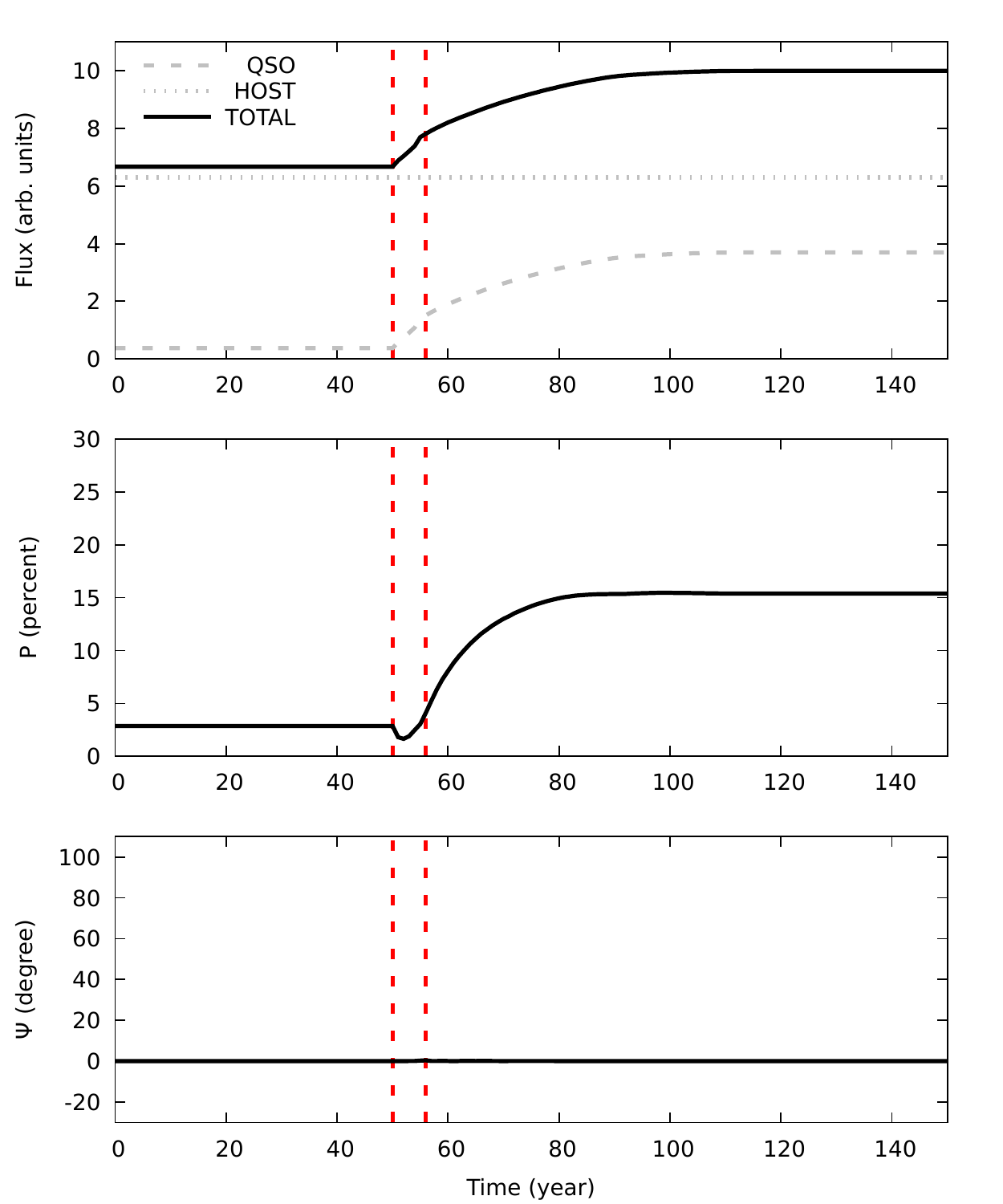}
        \caption{\small Observer inclination $\sim$ 60$^\circ$ 
          from the polar axis of the model.}    
        \label{Fig:60deg_brightening}
    \end{subfigure}
    \quad
    \begin{subfigure}[b]{0.475\textwidth}   
        \centering 
        \includegraphics[width=\textwidth]{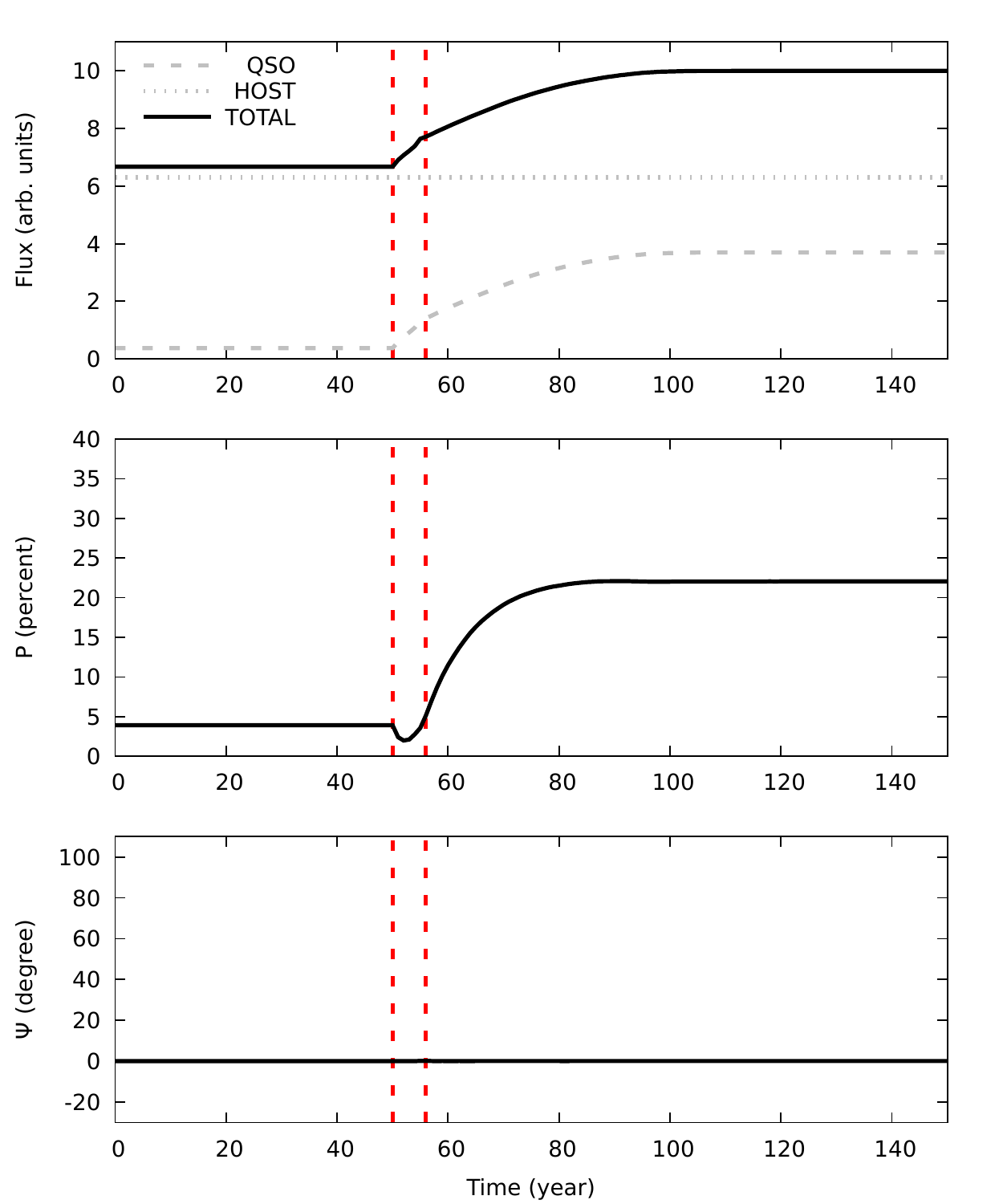}
        \caption{\small Observer inclination $\sim$ 80$^\circ$ 
          from the polar axis of the model.}  
        \label{Fig:80deg_brightening}
    \end{subfigure}
    \caption{Variation in normalized total flux (top panel), 
          polarization degree (middle), and polarization 
          position angle (bottom) from our changing-look 
          quasar model seen along four different viewing angles: 
          10$^\circ$ (\textit{a}), 45$^\circ$ (\textit{b}), 
          60$^\circ$ (\textit{c}), and 80$^\circ$ (\textit{d}). 
          In the total flux panels, we show the contribution
          of the quasar, the host galaxy, and the 
          combination of both using gray dashed lines 
          for the first two angles and a black solid line
          for the other two angles. Between the first and second 
          long-dashed vertical red lines, the central 
          engine brightened by a factor of 10 in 6 years.} 
    \label{Fig:QSO_brightening}
\end{figure*}

We now investigate the opposite trend in changing-look quasars: a brightening nucleus. We present our results in 
Figs.~\ref{Fig:10deg_brightening} and \ref{Fig:45deg_brightening} for type 1 objects and in Figs.~\ref{Fig:60deg_brightening} 
and \ref{Fig:80deg_brightening} for type 2 objects. The flux also varies by a factor 10 and the models are exactly the same.
In the following section we do not describe the total flux plots because they are the exact opposite to the dimming cases.
The polarization plots, however, show several different features.

In the case of type 1 objects (Figs.~\ref{Fig:10deg_brightening} and \ref{Fig:45deg_brightening}), the polarization degree
shows a rapid but small variation in $P$, while $\Psi$ remains constant. Most of the changes in $P$ occur during the 
brightening phase of the quasar and result in a partial decrease in polarization followed by a longer increase 
(a few years) and a smooth decrease with time. The sharp decrease in $P$ immediately follows the brightening of the central
engine and is easily explained by the large amount of unpolarized radiation that is created by the source. It takes less than a 
year for the brighter continuum flux to reach the equatorial scattering region, resulting in an enhanced production of polarized 
photons from that region. This is the reason for the subsequent increase in $P$ during the brightening phase. 
When the nucleus reaches a stable luminosity, the polarization stabilizes with a short delay due to the 
distance between the source and the reprocessing regions inside the nucleus. Because the brightening of the quasar
enhances the reprocessed polarized flux from the equatorial regions before the reprocessed polarized flux from the polar
winds, no variations in $\Psi$ are detected or expected. The main difference between the 10$^\circ$ and 45$^\circ$ 
inclinations is the value of $P$ (which remains low, below 1.5\% at best) and the time it takes for the 
polarization degree to stabilize. Similarly to the dimming cases, polarimetry can probe brightening phases, but the 
signatures are less prominent.

In the time-dependent polarization of type 2 changing-look quasars (Figs.~\ref{Fig:60deg_brightening} and 
\ref{Fig:80deg_brightening}), we observe a polarization degree that slightly decreases when the brightening of the central 
engine sets on, then a smooth and slow increase in $P$ for several decades, following the total flux. The polarization 
position angle remains constant and perpendicular to the projected symmetry axis of the system, as expected from type 2 
AGNs. The time-dependent behavior of $P$ shows in more detail that $P$ decreses during the 
first years to the enhanced production of equatorial, backscattered photons from the dust funnel opposite to the 
observer, as we discussed in the previous section. This stronger polarized flux, carrying a parallel
polarization angle, dilutes the observed polarization during the first years. The contribution of equatorial photons
then diminishes as the stronger polarized flux from the polar winds (situated at a larger radial distance from the 
source than the inner dust funnel) is visible above the circumnuclear dust horizon. The polarization then steadily increases
together with the total flux up to a constant and high ($>$10\%) value. The two cases (60$^\circ$ and 80$^\circ$)
are only discernible by their final polarization degree.

\subsubsection{Comparison with changing-look Seyferts}
\label{Model:Dynamic:Seyferts}

\begin{figure*}
    \centering
    \begin{subfigure}[b]{0.475\textwidth}
        \centering
        \includegraphics[width=\textwidth]{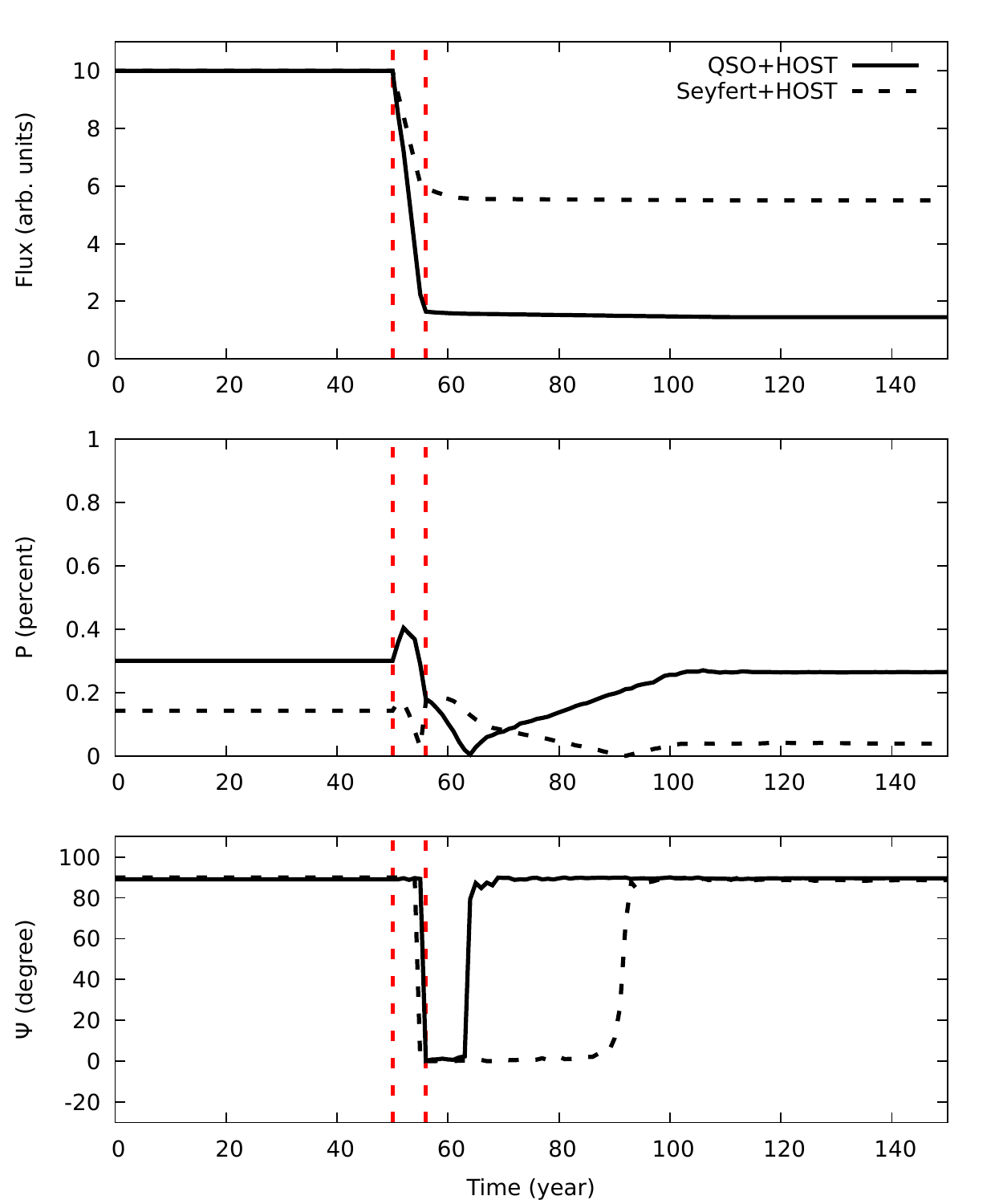}
        \caption{\small Observer inclination $\sim$ 10$^\circ$ 
          from the polar axis of the model.}    
        \label{Fig:Comp10deg}
    \end{subfigure}
    \hfill
    \begin{subfigure}[b]{0.475\textwidth}  
        \centering 
        \includegraphics[width=\textwidth]{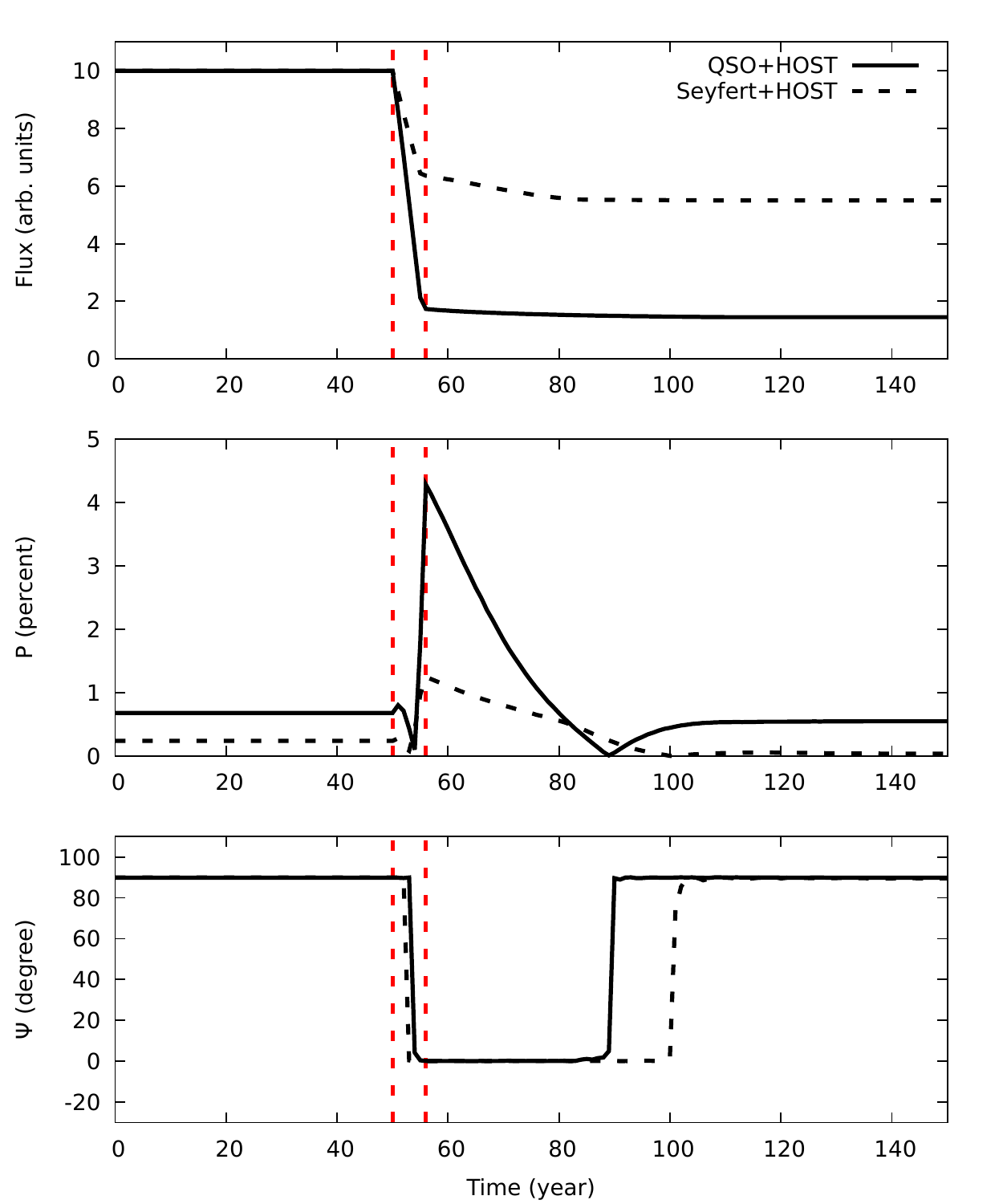}
        \caption{\small Observer inclination $\sim$ 45$^\circ$ 
          from the polar axis of the model.}     
        \label{Fig:Comp45deg}
    \end{subfigure}
    \vskip\baselineskip
    \begin{subfigure}[b]{0.475\textwidth}   
        \centering 
        \includegraphics[width=\textwidth]{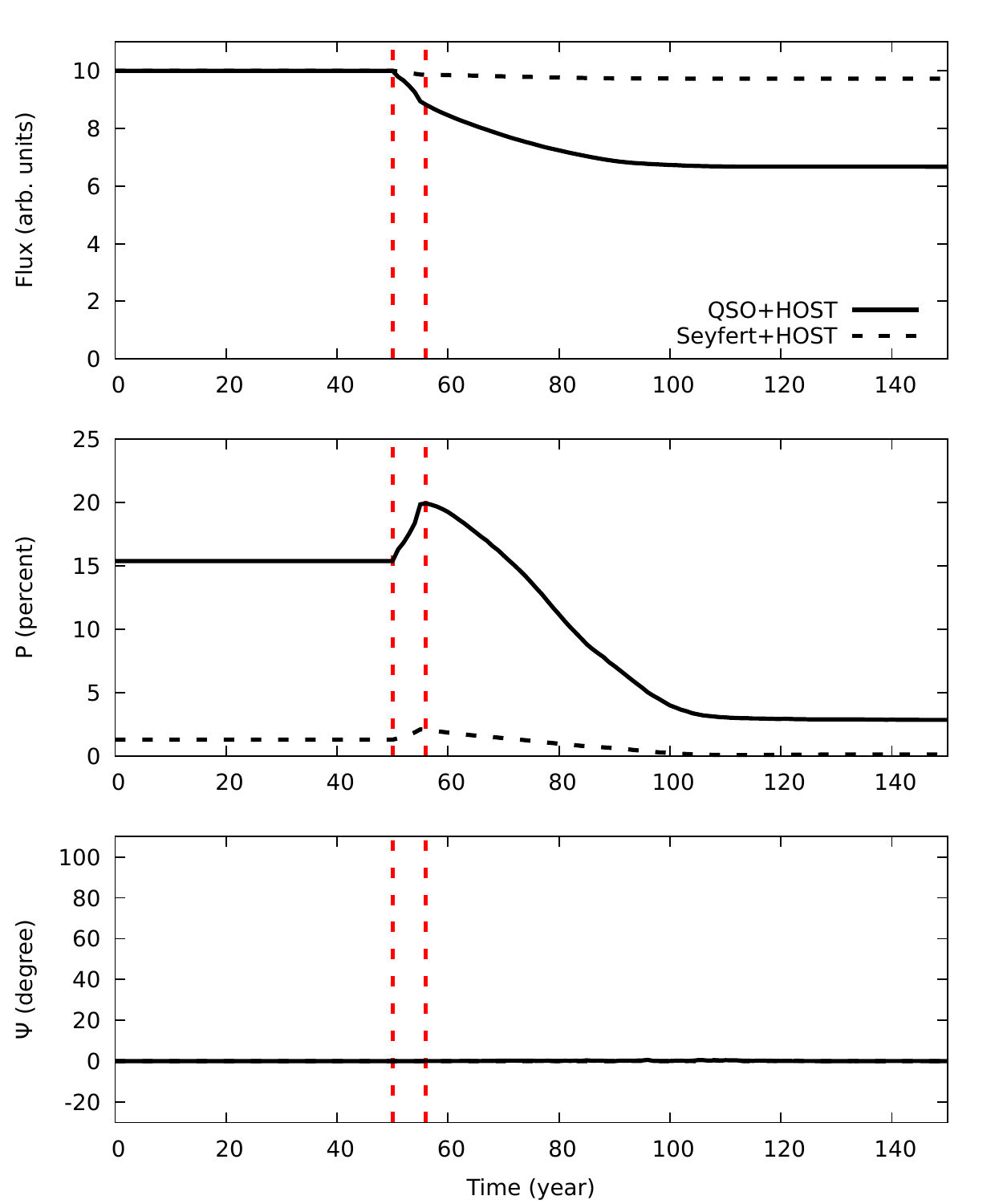}
        \caption{\small Observer inclination $\sim$ 60$^\circ$ 
          from the polar axis of the model.}    
        \label{Fig:Comp60deg}
    \end{subfigure}
    \quad
    \begin{subfigure}[b]{0.475\textwidth}   
        \centering 
        \includegraphics[width=\textwidth]{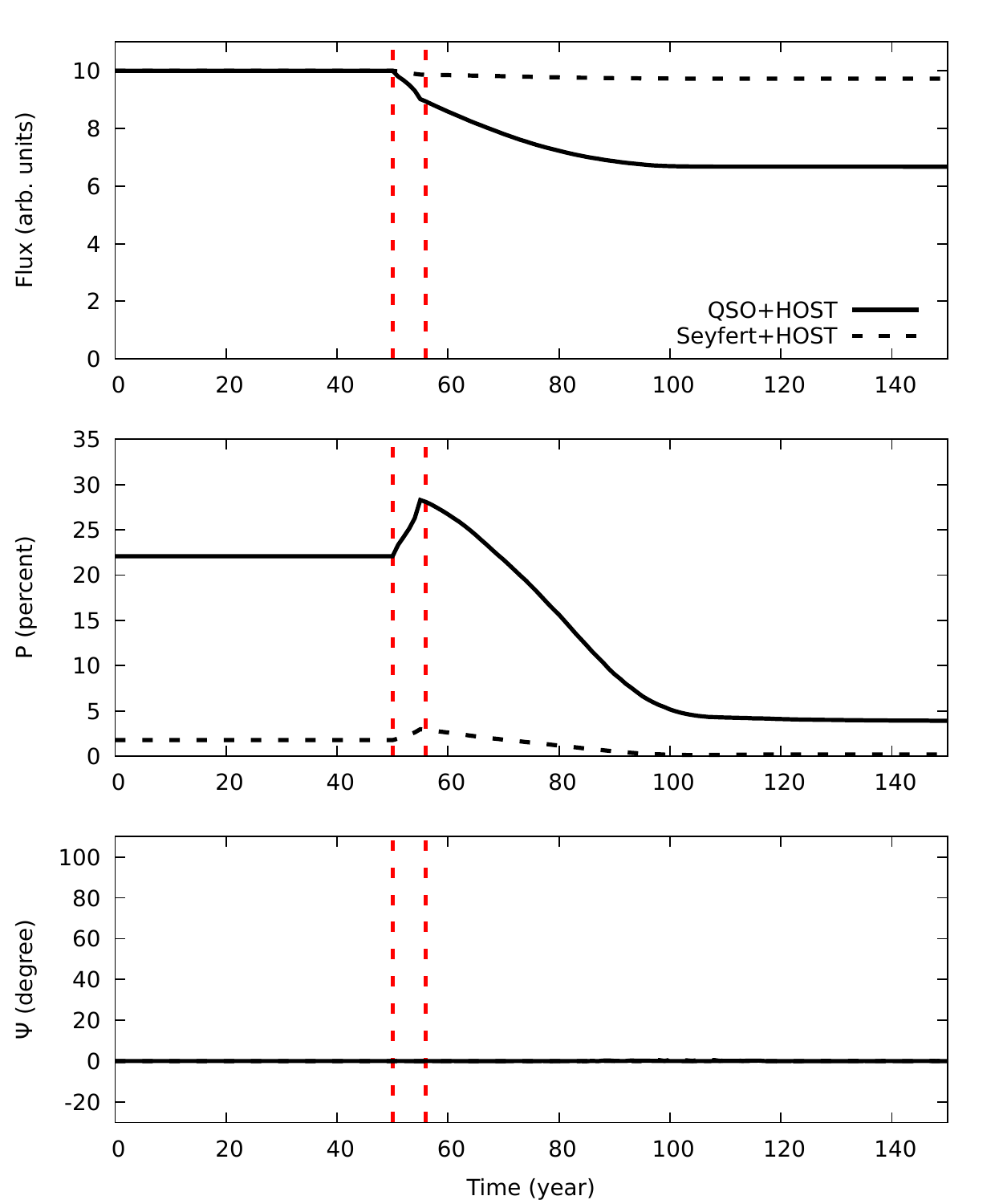}
        \caption{\small Observer inclination $\sim$ 80$^\circ$ 
          from the polar axis of the model.}  
        \label{Fig:Comp80deg}
    \end{subfigure}
    \caption{Comparison between the polarization time series 
          of changing-look quasars (solid line) and Seyfert 
          galaxies (dashed lines). Only the relative contribution 
          of the host galaxy to the total and polarized type 1 
          fluxes is different: 5\% in the quasar case, and 50\% for 
          Seyfert galaxies. The rest of the panel description 
          is similar to the caption of Fig.~\ref{Fig:QSO_dimming}.} 
    \label{Fig:Seyfert_dimming}
\end{figure*}

We finally investigate the case of Seyfert galaxies. We fixed the host contribution to 50\% of the total flux in 
type 1 views according to the average host contribution observed in nearby low-luminosity AGNs \citep{Kauffmann2003}.
When the nucleus is obscured at high inclinations (type 2 view), up to 97\% of the AGN radiation comes from the host galaxy.
The main result we obtained (see Fig~\ref{Fig:Seyfert_dimming}) is that the polarization time series of changing-look 
Seyfert galaxies are very similar to those of quasar, but are far more strongly diluted because starlight from the host galaxy is stronger. This is true for both dimming and brightening phase, therefore we only show the dimming case.
Fig~\ref{Fig:Seyfert_dimming} shows that the main difference in terms of polarization occurs 
at very low inclinations ($i \sim$ 10$^\circ$). At the onset of the dimming phase, the polarization 
degree experiences a sharp decrease that is associated with an orthogonal flip of the polarization angle. $P$ then rapidly 
increases before it returns to a stability period after a second flip of $\Psi$. This behavior has been observed for 
higher inclinations (see, e.g., the 45$^\circ$ case). We observe this in low-inclined changing-look 
Seyferts but not in low-inclined changing-look quasars because the host contribution is stronger. 
In both cases the polarization degree emerging from equatorial or polar scattering is low at $i \sim$ 10$^\circ$, 
but in quasars the central engine largely dominates the unpolarized flux from the host so that the polarization 
degree is less diluted. We also observe longer orthogonal flip periods of the polarization position angle for Seyfert 1 
inclinations. The previous simulations showed that the reversal in polarization angle takes place when 
the polarization of the polar echo becomes equal to the equatorial polarization. These two polarizations 
are fixed by the brightness of the central nucleus, which in the case of Seyfert galaxies is much lower than for
quasars. We observe the polar echo for a longer period because its polarized flux overwhelms the unpolarized flux from 
the host and the weaker polarized flux from the equatorial scattering region for a longer time. This explains the $\text{about}$ 15 years 
longer orthogonal flip of the polarization position angle in Fig~\ref{Fig:Comp45deg}. For type 2 viewing angles, there 
is no difference other than a stronger dilution of $P$ for lower luminosity AGNs. Our modeling 
shows that quasars are the best observational targets for monitoring fast and strong variations in polarization after
a change of look.

\section{Application to Mrk~1018}
\label{Application}

We showed that polarimetry is a powerful method for following the evolution of CLAGNs, in particular when 
the state transition is due to variations in mass accretion rate. Unfortunately, no multi-epoch 
polarimetric data are available so far for a quasar or a Seyfert galaxy that has shown a state transition that might be 
associated to intrinsic variation and is therefore likely to generate echoes. To support such eagerly awaited observations, 
we selected the best-documented case, the changing-look Seyfert galaxy Mrk~1018, and applied our numerical tool to
make quantitative predictions for a future observing campaign.

\begin{figure}
  \centering
  \includegraphics[trim = 0mm 0mm 0mm 5mm, clip, width=9.4cm]{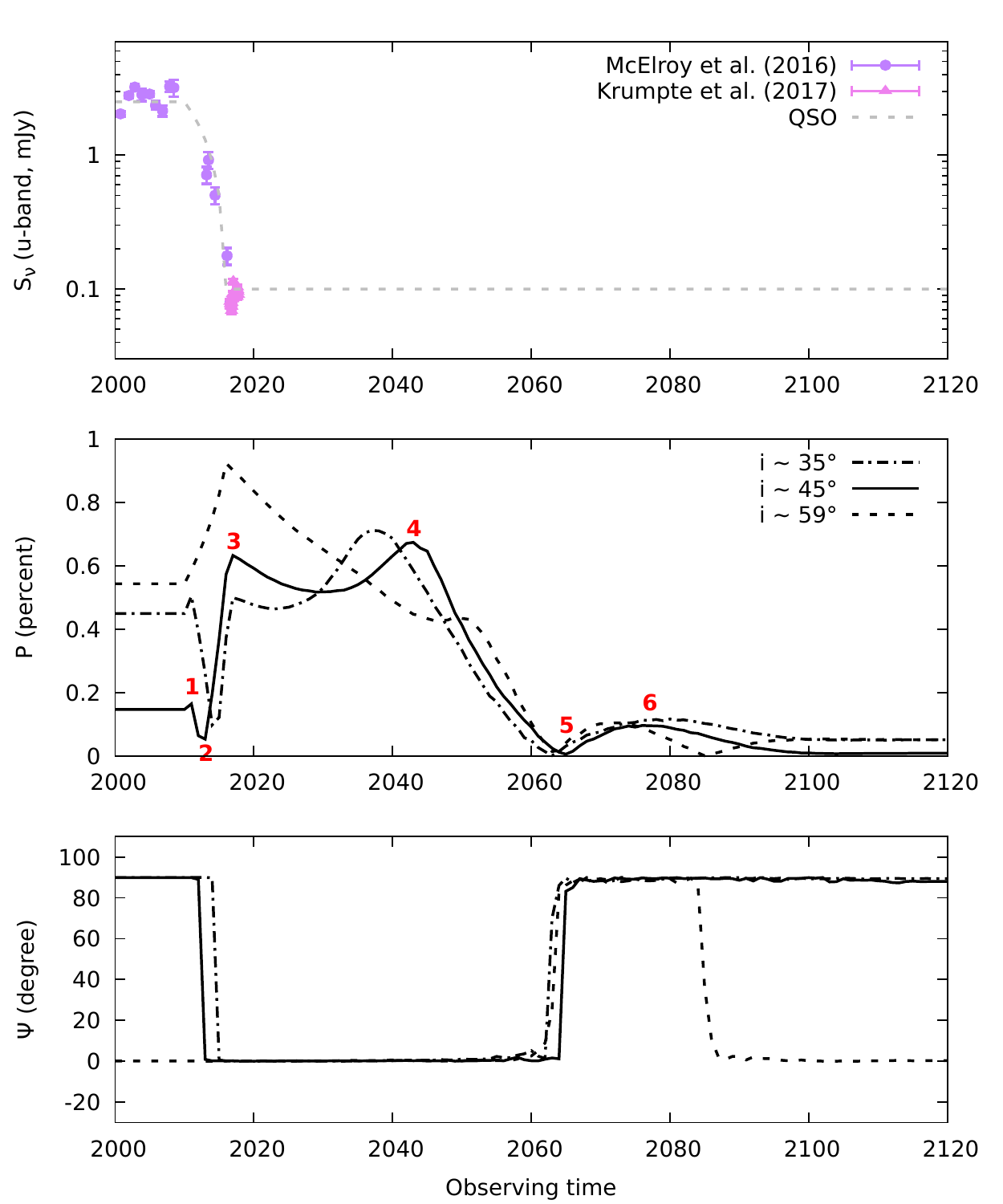}
  \caption{Top panel: Time series of archival u-band flux density S$_\nu$
           for the AGN in Mrk~1018 \citep{McElroy2016,Krumpe2017}. 
           The host galaxy contribution has been carefully 
           removed by the authors. We plot in gray the flux
           function we used in our simulation to mimic the 
           dimming of the AGN (host contribution removed).
           Middle and bottom panels: Model predictions for 
           the polarization degree and polarization angle for 
           different nuclear inclinations. See text for 
           an explanation of the red numbers.}
  \label{Fig:Mrk1018}
\end{figure}

Mrk~1018 is a type 1 AGN situated at a Hubble distance of 183.7~Mpc (heliocentric  redshift = 0.04244). It is a peculiar 
target because it is one of the few low-luminosity AGNs (Seyfert galaxy) known to have transitioned many times across the spectral 
sequence. Mrk~1018 varied from type 1.9 to 1 between 1979 September and 1984 January \citep{Cohen1986},
then returned again to type 1.9 in 2015 \citep{McElroy2016}. The AGN is currently stabilizing at low nuclear optical fluxes 
since 2017 \citep{Krumpe2017}, as seen in Fig.~\ref{Fig:Mrk1018} (top panel). These changes of look are explained by a variation
in the accretion rate onto the SMBH \citep{Husemann2016}.

The very recent state transition of Mrk~1018 makes it an excellent and priority target for polarimetric observations in order
to assess the origin of the change of  look. Unfortunately, only one polarimetric observation of Mrk~1018 has been made so far. It was made on 9 February 1986.
At that time, \citet{Goodrich1989} reported a polarization degree of 0.28 $\pm$ 0.05\% in the 4180 -- 6903~\AA~band. 
It is difficult to use this measurement because the reported polarization degree is very low; this might
  be due to interstellar polarization from the host galaxy or to foreground dust in our Galaxy. The associated polarization
position angle ($\theta$ = 165.1 $\pm$ 5.2$^\circ$) is unlikely to be meaningful either becaue we do not know the origin of the 
polarization. In addition, we do not know the position angle of the (sub)parsec scale radio 
structure in Mrk~1018 either. Milliarcsecond angular resolution radio 
imaging of Mrk~1018 with the Very Low Baseline Array (VLBA) does not resolve the core-jet structure of the AGN (P\'erez-Torres, private communication).
We therefore do not know if Mrk~1018 had a parallel or perpendicular polarization angle in 1986. New polarimetric measurements
of $\theta$ might indicate whether the polarization angle has rotated since then. 

We modeled Mrk~1018 using the same general AGN geometry as in Sect.~\ref{Model}. We fixed the host galaxy contribution
during the bright (pre-2011) and dim (post-2016) states of Mrk~1018 using the luminosity ratio of the nucleus to the host 
galaxy as a function of epoch presented in \citet{Kim2018}. The authors analyzed multi-epoch Sloan Digital Sky Survey (SDSS) g-band and SWIFT B-band 
imaging data and derived a luminosity fraction L$_{\rm AGN}$/L$_{\rm host}$ $\sim$ 0.4 in the bright state and 
$\sim$ 0.03 in the dim state. We allowed the nucleus to decreases its flux by a factor 25, such as observed by 
\citet{McElroy2016}. Finally, we observed the system along a specific range of viewing angles. According to \citet{Walton2013},
the AGN inclination is 45$^{+14}_{-10}$ degrees. This value was found using reflection-based modeling of 
the broad-band X-ray spectrum of Mrk~1018 and remains the only estimate for Mrk~1018 so far \citep{Marin2016}. We 
therefore investigate in Fig.~\ref{Fig:Mrk1018} three inclinations: 35$^\circ$, 45$^\circ$ , and 59$^\circ$, which covers the 
entire inclination range derived by \citet{Walton2013}. We note that at $i$ = 59$^\circ$, the line of sight crosses
the equatorial dust. This would make Mrk~1018 a type-2 Seyfert galaxy at all epochs but, for the academic interest of the study, we 
also consider this inclination.

\begin{figure}
  \centering
  \includegraphics[trim = 0mm 0mm 0mm 0mm, clip, width=9.0cm]{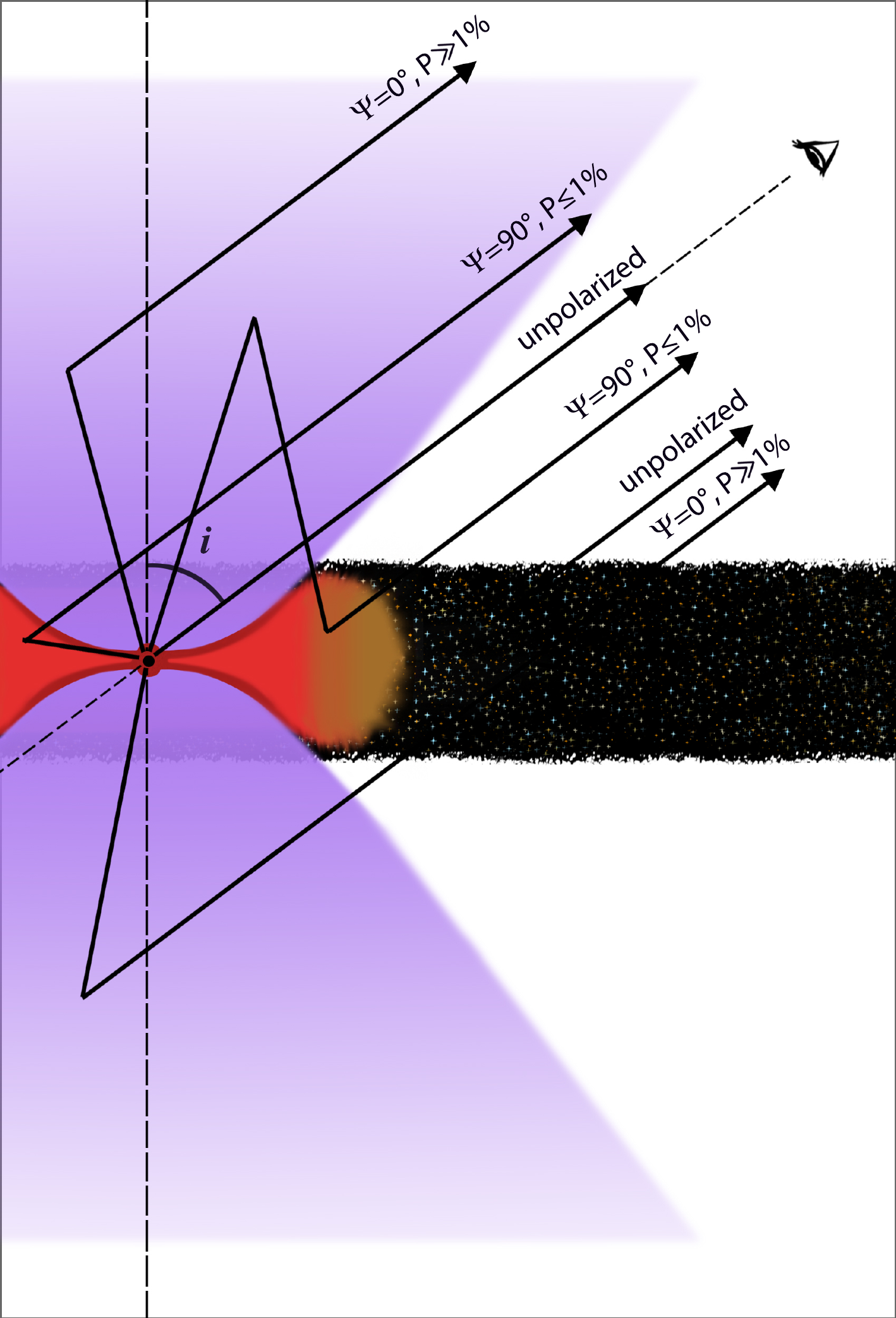}
  \caption{Illustration of the various paths
          taken by the photons for a typical
          type 1 viewing angle ($i \sim$ 45$^\circ$).
          The expected polarization signature of
          emission or reprocessing regions is indicated.
          The breaks in the light rays 
          are due to scattering onto electrons or 
          dust grains. See text for details.}
  \label{Fig:Scheme_with_arrows}
\end{figure}

We show in the top panel of Fig.~\ref{Fig:Mrk1018} the observed time-dependent u-band flux density S$_\nu$ of Mrk~1018 
(in mJy). The source flux function we used was set to become fainter by a factor of 25 between 2011 and 2017 and is shown in gray.
We allowed the system to reach a stable configuration following 2017 for the next century but we note that a new state transition might very well occur in the future. For simplicity, we only show one transition here. In the middle and bottom panels of 
Fig.~\ref{Fig:Mrk1018} we show the polarization degree and polarization position angle, respectively. For all inclinations,
the polarization degree appears low, below 1\% because of the strong dilution by the host galaxy. These polarization levels are 
in good agreement with the unique measurement made by \citet{Goodrich1989}. The three different inclinations are displayed
using different line types. The polarization behavior is more complex than what was shown in the 
previous sections. The main reason for this increasing complexity is the stronger variation in the core flux (a factor
25 instead of the factor 10 explored in the previous sections) that will generate secondary echoes. To better understand the 
various polarization signatures we expect to observe, we report the various possible trajectories of photons in 
Fig.~\ref{Fig:Scheme_with_arrows} and number the principal features in red in Fig.~\ref{Fig:Mrk1018} (middle panel) 
for the $\sim$ 45$^\circ$ inclination. We give an explanation of each feature below.

\begin{enumerate}
  \item When the dimming sets on, we observe an almost negligible increase in $P$. This feature has 
        been explained for Figs.~\ref{Fig:10deg} and \ref{Fig:45deg} and corresponds to the delayed
        response of the equatorial scattering region to the nuclear dimming.
  \item The local dip in $P$ is due to the orthogonal flip of $\Psi$. The superposition 
        of parallel polarization from the equatorial regions and perpendicular polarization 
        from light echoes in the polar winds cancel each other out.
  \item The maximum of perpendicular polarization is due to the propagation of the bright-light echo inside 
        the upper part of the biconical outflow. At this point, the polarized echo starts to fade away.
  \item The second maximum of the perpendicular polarization is due to the propagation of the bright-light echo inside 
        the lower part of the biconical outflow. The delay between the two polar echoes is a 
        geometric effect that is due to the inclination of the AGN. The second light echo then starts to fade away as well.
  \item The second dip in $P$ is due to the second orthogonal flip of $\Psi$. The dominance of polar-light 
        echoes ends and the observed polarization again becomes parallel.
  \item A final bump in polarization is detected before $P$ resumes a stable level (which is highly diluted
        by the strong starlight from the host). This bump is due to multiple scattering of photons from
        the light echo in the polar regions that traveled back to the AGN core and scattered onto the
        external surfaces of the circumnuclear dust \citep{Marin2012,Grosset2018}.
\end{enumerate}

The same reasoning holds for the 35$^\circ$ inclination. In the case of $i$ = 59$^\circ$, the change in polarization is similar 
to what was described in Figs.~\ref{Fig:60deg} and \ref{Fig:80deg}. In addition, a secondary polarized echo from the lower part
of the biconical structure is visible in about 2050. The scattering of perpendicularly polarized photons from the polar winds on the 
external surfaces of the circumnuclear dust is highlighted by a new flip of the polarization angle in about 2085. 
Interestingly, Figs.~\ref{Fig:QSO_dimming} and \ref{Fig:QSO_brightening}  the polarization features associated with
the propagation of the  light echo inside 
the lower part of the biconical outflows are absent.
Using {\sc stokes}-computed polarization maps, \citet{Marin2012} showed that most of the observed polar flux is due
to scattering within the first parsecs of the biconical outflow. At type 1 inclinations, the base of the lower part of the
biconical outflow is hidden by the distribution of equatorial dust. The photon flux from the southern wind base is therefore mainly
absorbed by the optically thick dusty torus, and radiation scattered at larger distances is relatively weak.
However, if the flux variation is extreme, the light echoes are stronger and more photons can reach more
distant parts of the lower biconical outflow that are not hidden by the thick dusty torus. This stronger and delayed echo of polarized
flux explains why we detect an additional polarization signature when the continuum source has dimmed by a very large 
factor. 

Ultimately, we demonstrated that a polarimetric campaign targeting Mrk~1018 over the next decade could tell us much 
about the physics and morphology of the innermost components of nearby AGNs. Even if the polarization is strongly diluted by the host starlight,
polarization variability can be measured independently of the contamination, provided that the instrumentation is accurate enough.
We discuss this point in more detail in Sect.~\ref{Discussion:Degeneracies}.

\section{Discussion}
\label{Discussion}
Our modeling has demonstrated that changing-look AGNs are expected to present strong and long polarized echoes that are associated with both 
a dimming or brightening phase. While the total flux variation lasts for only a few years, the polarized echo lasts for decades. This
increases our chances of detecting and characterizing the accretion processes and the geometry of the various reprocessing regions 
in quasars and Seyfert galaxies. In Sect.~\ref{Discussion:Degeneracies} we discuss the observational constraints we can place 
on nearby and distant AGNs while estimating the relative importance of model degeneracies. We also report in Sect.~\ref{Discussion:Observations}
the most interesting cases for future polarimetric observations in order to better understand the underlying causes of the 
accretion rate variability that is responsible for CLAGNs.

\subsection{Observational constraints and model degeneracies}
\label{Discussion:Degeneracies}
After the onset of the dimming phase in type 1 CLAGNs, a slight increase in $P$ is followed by a strong decrease 
that is associated with an orthogonal flip of the polarization position angle (see Figs.~\ref{Fig:10deg} and \ref{Fig:45deg}). 
As we described, the increase in $P$ is due to the continuum echoing onto the equatorial scattering region (parallel $\Psi$), which is soon 
substituted by the polarized echo on the polar winds (perpendicular $\Psi$). A triggered polarization monitoring of a CLAGN
currently undergoing a state transition from type 1 to type 2 may very well measure the delay between the dimming 
and the light echo from the equatorial scattering region and/or the polar structure. This time information could then be translated directly
into geometric constraints in the form of upper limits on the inner radius of the two reprocessing regions. The inferred inner 
radii would be dependent on the optical depth of the specific scattering region, but we already know that the polar winds 
and the equatorial scattering region are not highly opaque \citep{Osterbrock1991}. In addition, the orthogonal flip of $\Psi$ would clearly indicate
which scattering region dominates in terms of flux. Dilution by the host does  not interfere with these observational 
constraints (see, Fig.~\ref{Fig:Comp45deg}), it can only make the observation more time-consuming because of the lower 
expected polarization degree in Seyfert galaxies. The duration of the orthogonal flip in polarization angle is directly 
related to the inclination of the nuclear system. The longer the flip, the more inclined the object. However, 
this needs to be carefully modeled. Fig.~\ref{Fig:Comp10deg} clearly demonstrates that a high host or AGN flux ratio may also
increase the duration of the $\Psi$ reversal. In such cases, it will be mandatory to determine 
the host contribution before the true nuclear inclination of the AGN is determined. This means that a new and 
spectroscopically independent measurement of the global AGN inclination can be achieved in CLAGNs. If no orthogonal flip of 
$\Psi$ is observed, it means that the system is seen along an equatorial viewing angle, that is, the observer's line of sight crosses the circumnuclear dust region. The statistical counting of CLAGNs with and without polarization angle reversals
would bring crucial information on the average half-opening angle of the dusty torus and help determine its formation 
process. If the torus results from mass transfer from the host to the AGN, its 
geometry is expected to be flatter than when the torus is in fact a wind that flows out at low velocities from the AGN to 
the host (see, e.g., \citealt{Krolik1988,Konigl1994,Lovelace1998,Czerny2011,Dorodnitsyn2012}). Moreover, because CLAGNs 
are detected up to redshifts higher than 0.8 \citep{Graham2019}, it will be possible to determine whether the geometric 
configuration of the torus evolves with redshift and luminosity. 

A final remark must be added concerning model degeneracies in our analysis. The features detected 
in the polarization time series are model dependent. As we mentioned, the AGN or host flux ratio plays a role, but 
this may be counterbalanced by dedicated measurements. The geometry and optical depth of 
the AGN reprocessing components will mainly affect the delays seen in polarized echoes. They do not play a major role 
in the observed degrees of polarization, at least for type 1 objects \citep{Marin2012}. The difference is more 
noticeable for equatorial lines of sight, where the optical depth and half-opening angle of the polar winds will 
provide most of the polarized flux. Fortunately, the extended polar winds are not too difficult to 
resolve spatially, at least in nearby AGNs. From narrow-line emission intensity and widths, constraints on the 
electron optical depth can be derived while medium-resolution intensity [O III] maps and long-slit spectra can provide 
an insight into the wind half-opening angle. \citet{Fischer2013,Fischer2014} used this technique to determine the inclination of the extended biconical outflows with respect to the plane of the sky with a 
5$^\circ$ precision. The main 
source of degeneracies remains the spatially unresolved innermost AGN components. Although interferometry is now able 
to probe the subparsec-scale regions in the most luminous and closest AGNs \citep{Gravity2018}, such observations
will remain impossible for the vast majority of AGNs for the foreseeable future. Careful modeling of observational 
(spectro)polarimetric and photometric data will help to reduce the number of free parameters, but the uncertainties 
associated with the method remain to be tested.

\subsection{Past polarimetric observations of CLAGNs}
\label{Discussion:Observations}
In the case of changing-look quasars, two publications currently report the measurement of associated optical polarization: 
\citet{Hutsemekers2017} and \citet{Hutsemekers2019}. The first paper only considered one CLAGN, J101152.98+544206.4, a changing-look
quasar ($z$ = 0.246) discovered by \citet{Runnoe2016}, which transitioned from a bright state in 2002 to a faint state in 2017. The 
2017 optical linear continuum polarization degree of this object is 0.15 $\pm$ 0.22\%, consistent with null polarization, therefore it was not possible to estimate its
polarization angle. The very weak (most probably null) polarization of J101152.98+544206.4 tells us about the 
probable inclination of this quasar (close to pole-on, \citealt{Marin2017}). According to our simulations, the polarized-light echo 
of the state transition is currently propagating, but because of the very low expected polarization degree (see Fig.~\ref{Fig:10deg}), detecting polarization variations is difficult. The second paper presents a more promising 
source. As shown in \citet{Hutsemekers2019}, polarization echoes can explain the high polarization ($P$ = 6.8\%, 
$\theta$ = 71 $\pm$ 3$^\circ$) observed in the changing-look quasar J022652.24$-$003916.5 ($z$ = 0.625) that is presently in 
a faint state, assuming that this quasar is seen at intermediate inclination. Our simulations (see Fig.~\ref{Fig:45deg}) show 
that a regular decrease in polarization degree is expected during the next decades, ultimately with a reversal of the 
polarization angle. Because of the high polarization degree of the object, a monitoring campaign could easily prove or disprove 
the accretion rate variation scenario and narrow down the real inclination of the quasar, depending on the exact date
of the expected orhtogoanl flip of the polarization position angle. In the same publication, the low polarization degrees ($\le$ 1.10\%) 
measured for the other changing-look quasars in a type 2 state are also compatible with the existence of polarization echoes 
if these objects are seen at low inclinations. Variations are expected, but lie below the accuracy of available measurements. 
In these objects, a reversal of the polarization angle is probably the most sensitive way to unveil to presence of 
polarization echoes.

In Seyfert galaxies, changes of look are explained by variation in accretion rate in at least three of them: Mrk~1018 \citep{Husemann2016}, 
Mrk~590 \citep{Denney2014}, and NGC~2617 \citep{Shappee2014}). Polarization echoes can thus be reasonably expected. Unfortunately, almost no archival polarization data are available for these objects. As we mentioned, the polarization of Mrk~1018 was measured
in 1986 by \citet{Goodrich1989} when the AGN was in a type 1 state: $P$ = 0.28 $\pm$ 0.05\%, and $\theta$ = 165.1 $\pm$ 5.2$^\circ$. 
The polarization of Mrk~590 was measured in 1976 by \citet{Martin1983} when the AGN was in a type 1 state: $P$ = 0.32 $\pm$ 0.30\%, and 
$\theta$ = 105.9 $\pm$ 26.6$^\circ$. However, this measurement is unlikely to be useful because the reported polarization 
degree $P$ is on the same order as $\sigma_{\rm P}$ so that the polarization angle is undefined 
($P$ is a positive quantity with non-Gaussian errors; \citealt{Simmons1985}). Finally, the polarization of NGC~2617 was measured in 1998 by 
\citet{Wills2011} when the AGN was in a type 2 state: $P$ = 0.43 $\pm$ 0.15\%, and $\theta$ = 16 $\pm$ 10$^\circ$. In the two 
later cases, if the observed polarization is really due to scattering of nuclear light, the low polarization degrees 
might indicate that the AGNs are seen at low inclinations, but strong dilution by the host can also be expected. 
Estimating the inclination of AGNs is complex \citep{Marin2016}, and the values we have for Mrk~1018 (45$^{+14}_{-10}$, 
\citealt{Walton2013}) and Mrk~590 (17.8$^{+6.1}_{-5.9}$, \citealt{Wu2001}) might very well be revised based on a polarization 
monitoring of those objects. Overall, there has been no new polarization measurement of any 
changing-look Seyfert galaxies since the end of the 1990s, while the phenomenon of type transition has become commonly known. 
This fact demands new polarization measurements of all CLAGNs in order to create a database for future analyses.

\section{Conclusions}
\label{Conclusions}
We have undertaken a series of single-wavelength simulations to predict the polarization time variations of CLAGNs. We investigated both the 
low-luminosity nearby cases (Seyferts) and the high-luminosity distant ones (quasars). We simulated a dimming or a 
brightening scenario of the central engine by several times in few years and observed the propagation of polarized-light 
echoes on timescales that are much longer than what photometry alone can reveal. The time-dependent polarization 
signatures, both in terms of polarization degree and angle, highlight the different AGN components at various scales 
(from subparsec to kiloparsec regions). They allow us to place constraints on the inner radii and optical depth of the 
various reprocessing components. An orthogonal flip of the polarization position angle in type 1 CLAGNs provides a new
method for estimating the inclination of the nucleus, independently of the associated spectroscopic signatures.
We applied our simulations to a specific case, Mrk~1018, which is one of the best-documented CLAGN 
cases. We predict its time-dependent polarization signatures after the recent dimming by a factor 25 of the AGN flux. 
Our numerical analysis demonstrates that crucial information on the innermost structure and on the inclination angle
of the entire system may be gained by implementing a new campaign of polarization monitoring of CLAGNs. In 
particular, the spectropolarimetric mode of large-class telescopes would greatly improve our understanding of the 
physical processes that are responsible for the dimming or brightening of CLAGNs. A direct comparison between the polarization
spectra of the nucleus and of the large-scale polar region would help us to follow the evolution of the light echo.
Such a program would help constrain the scenario of the mass accretion rate variation and place strong upper limits on the inner radii of 
the equatorial and polar scattering regions even in spatially unresolved objects. In addition, new polarization measurements of 
CLAGNs are crucial for determining whether unification by orientation is 
still dominant in AGN models. We therefore strongly advocate a renewed observational effort to fully characterize the spectroscopic and imaging
polarization properties of CLAGNs.

\begin{acknowledgements}
The authors would like to thank Robert ``Ski'' Antonucci for refereeing this paper and bringing many insightful comments.
The authors are also  grateful to Miguel \'A. P\'erez-Torres for discussing his private VLBA and EVN observations of 
Mrk~1018. FM would like to thank the Centre national d'\'etudes spatiales (CNES) who funded his post-doctoral grant 
``Probing the geometry and physics of active galactic nuclei with ultraviolet and X-ray polarized radiative
transfer''. DH is senior research associate F.R.S.-FNRS. Finally, the authors are grateful to Jules Garreau (\textcolor{cyan} 
{jul.garreau@wanadoo.fr}) for his artworks of the AGN model.
\end{acknowledgements}

\bibliographystyle{aa}
\bibliography{biblio}

\end{document}